\documentclass[12pt,preprint,eqsecnum]{aastex}

\usepackage{emulateapj5}

\usepackage{graphicx}
\usepackage{dcolumn}
\usepackage{bm}
\usepackage{epsf}

\lefthead{BAUMGARTE \& SHAPIRO}
\righthead{COLLAPSE OF A MAGNETIZED STAR TO A BLACK HOLE}

\begin{document}

\title{Collapse of a Magnetized Star to a Black Hole}

\author{Thomas W. Baumgarte\altaffilmark{1,2} and 
	Stuart L. Shapiro\altaffilmark{2,3}}

\affil{\altaffilmark{1}
	Department of Physics and Astronomy, Bowdoin College,
	Brunswick, ME 04011}

\affil{\altaffilmark{2}
	Department of Physics, University of Illinois at
	Urbana-Champaign, Urbana, Il~61801}

\affil{\altaffilmark{3}
	Department of Astronomy and NCSA, University of Illinois at
     	Urbana-Champaign, Urbana, Il~61801}

\begin{abstract}
We study of the collapse of a magnetized spherical star to a black
hole in general relativity theory. The matter and gravitational fields
are described by the exact Oppenheimer-Snyder solution for the
collapse of a spherical, homogeneous dust ball.  We adopt a
``dynamical Cowling approximation'' whereby the matter and the
geometry (metric), while highly dynamical, are unaffected by the
electromagnetic fields.  The matter is assumed to be perfectly
conducting and threaded by a dipole magnetic field at the onset of
collapse.  We determine the subsequent evolution of the magnetic and
electric fields without approximation; the fields are determined
analytically in the matter interior and numerically in the vacuum
exterior. We apply junction conditions to match the electromagnetic
fields across the stellar surface.  We use this model to experiment
with several coordinate gauge choices for handling spacetime evolution
characterized by the formation of a black hole and the associated
appearance of singularities. These choices range from ``singularity
avoiding'' time coordinates to ``horizon penetrating'' time
coordinates accompanied by black hole excision.  The later choice
enables us to integrate the electromagnetic fields arbitrarily far
into the future. At late times the longitudinal magnetic field in the
exterior has been transformed into a transverse electromagnetic wave;
part of the electromagnetic radiation is captured by the hole and the
rest propagates outward to large distances.  The solution we present
for our simple scenario can be used to test codes designed to treat
more general evolutions of relativistic MHD fluids flowing in strong
gravitational fields in dynamical spacetimes.
\end{abstract}

\maketitle

\section{Introduction}
\label{sec1}

Magnetic fields play a crucial role in determining the evolution of
many relativistic objects.  In a companion paper (Baumgarte \&
Shapiro, 2002, hereafter Paper I) we assembled the complete set of
Maxwell--Einstein--magnetohydrodynamic (MHD) equations that
determine the self-consistent evolution of a relativistic, ideal MHD
gas in a dynamical spacetime. Our goal was to set down a formulation
of the equations that is suitable for numerical integration in full
$3+1$ dimensions.

In this paper (hereafter Paper II) we use these general relativistic
MHD equations to follow the gravitational collapse of a magnetized
star to a black hole. We solve this problem in a ``dynamical Cowling
approximation'', whereby the matter and gravitational field are given
by the Oppenheimer-Sndyer solution (Oppenheimer \& Snyder 1939) for
the collapse from rest of a spherical dust ball to a Schwarzschild
black hole. We assume that initially the star is threaded by a dipole
magnetic field and determine the subsequent evolution of the magnetic
and electric fields with time, both in the matter interior and vacuum
exterior.

We are motivated to tackle this problem partly to acclimate ourselves
to the task of numerically integrating the coupled
Maxwell-Einstein-relativistic MHD equations for a strong-field,
dynamical spacetime.  Though the background geometry is entirely
determined in this example, the spacetime is highly dynamical and
characterized by a very strong (black hole) gravitational field. The
goal is to solve for the electromagnetic fields to arbitrary late
times everywhere in space.  The restricted problem we pose here is
sufficiently rich to allow a comparison of coordinate gauge choices
for handling collapse to a black hole and dealing with the
accompanying spacetime singularity. Such gauge choices range from
``singularity avoiding'' time coordinates to horizon penetrating time
slices which allow for black hole excision. We experiment with these
different choices below.

The collapse of a magnetized star to a black hole is an
astrophysically important problem that has been discussed previously.
Some of the earliest treatments were based on a quasi-static approach
(Ginzburg \& Ozernoy 1965; Anderson \& Cohen 1970; Zel'dovich \&
Novikov 1971) which was demonstrated to give the incorrect asymptotic
decay of the fields with time (Price 1972a,b; see also Thorne
1971). The general problem of magnetic star collapse to a black role
obviously requires a detailed numerical treatment for solution of the
coupled Maxwell -- Einstein -- relativistic MHD system. Only now is
such a treatment feasible. In fact, we are unaware of any previous
analysis that determines the complete evolution of both the interior
{\it and} exterior electromagnetic fields during stellar collapse to a
black hole, even for the restricted spherical collapse problem posed
here.  [In a pioneering paper, Wilson 1975 used relativistic MHD to
follow the collapse of the interior of a magnetized star, but treated
the gravitational field as quasi-stationary].  Our solution should
serve as a useful testbed for designing codes capable of handling more
complicated scenarios involving relativistic MHD gravitational
collapse. Such scenarios may play a crucial role in determining the
outcome of stellar core collapse in a supernova, the fate of a
hypermassive, differentially rotating remnant of a binary neutron star
merger (Baumgarte, Shapiro \& Shibata 2000; Shapiro 2000), the
generation of gamma-ray bursts via the collapse of rotating massive
stars to black holes (MacFadyen \& Woosley 1999), and the collapse of
supermassive stars to supermassive black holes (see, e.g., Rees 1984,
Baumgarte \& Shapiro 1999, New \& Shapiro 2001, and references
therein.)

This paper is partitioned as follows: In Section \ref{sec2} we present
a brief overview of magnetized Oppenheimer-Snyder collapse, and in
Section \ref{sec3} we provide a review of the matter and metric
fields.  In Section \ref{sec4} we derive analytic Newtonian and
relativistic solutions for the interior electromagnetic fields, and in
Section \ref{sec5} we set up Maxwell's equations and boundary
conditions for determining the exterior fields.  In Section \ref{sec6}
we provide electrodynamic initial data, both for the interior and the
exterior.  In Section \ref{sec7} we present numerical results for the
evolution of the exterior fields in three different time slices.  In
particular, in Section \ref{sec7.4} we present results in Kerr-Schild
coordinates, which allow us to follow the evolution to arbitrary late
times and to track the late-time fall-off of the fields.  We briefly
summarize our analysis in Section \ref{sec8}.  We also provide two
Appendices.  Appendix \ref{secA} derives a global electromagnetic
energy conservation criterion, and Appendix \ref{secB} contains the
finite-difference equations used in our simulations.

\section{Magnetized Oppenheimer-Snyder Collapse: Overview}
\label{sec2}

In this paper we consider the ``simplest'' case of magnetized stellar
collapse to a black hole: the collapse of a spherically symmetric,
homogeneous dustball, momentarily at rest at $t=0$. The matter and
gravitational fields are described by the exact Oppenheimer-Snyder
solution.

We solve for the electrodynamic fields in a ``dynamical Cowling
approximation'': the background geometry is dynamical and determined
by the imploding matter field alone; the matter and the geometry
(metric) are unaffected by the electrodynamic fields, which are
evolved in the unperturbed background.  Our approximation is physically
applicable whenever the magnetic field is sufficiently weak that it
satisfies the condition $B^2/8\pi \ll GM\rho/R$, where $G$ is the
gravitational constant, $\rho$ is the matter density and $M$ and $R$
are the mass and radius of the star. Note that for a frozen-in
B-field, both terms scale the same way with radius (i.e. $R^{-4}$)
during the collapse, so if the inequality is satisfied initially, it
is satisfied at all later times.

We seek the relativistic generalization of the solution for the
electromagnetic field generated inside and outside a collapsing,
homogeneous, perfectly conducting magnetized dust sphere which at
$t=0$ is threaded by a constant interior B-field, matched onto an
exterior magnetic dipole field. The solution to this problem requires
a numerical integration of Maxwell's equations in a dynamical
spacetime in which a black hole forms. While various aspects of this
problem have been treated previously, including the asymptotic field
behavior at late times in a static vacuum Schwarzschild spacetime, the
full dynamical solution has never been presented.

We will divide our time-dependent solution of the electrodynamic
fields into two domains: the matter interior solution (Section
\ref{sec4}) and vacuum exterior solution (Section \ref{sec5}). The
solutions are joined at the stellar surface by applying the junction
conditions for an electromagnetic field across a moving
discontinuity. Hereafter we adopt gravitational units whereby $G=1=c$.

\section{The Matter and Metric Fields: The OS Solution}
\label{sec3}

The Oppenheimer-Snyder solution (Oppenheimer \& Snyder 1939) can be
derived from the Euler equations, the continuity equation and the
Einstein field equations. Each fluid element follows a radial geodesic
in this approximation.  The interior metric is given by the familiar
(closed Friedmann) line element
\begin{equation} \label{int_metric}
ds^2=-d\tau^2+A^2(d\chi^2 +\sin^2\chi\,d\Omega^2).
\end{equation}
Here $\tau$ is the (proper) time coordinate, measured from the onset
of collapse, $\chi$ is a (Lagrangian) radial coordinate and $A$ is
related to $\tau$ through the conformal time parameter $\eta$,
\begin{eqnarray} 
A & = & {1\over 2} A_m(1+\cos\eta) \\
\tau & = & {1\over 2} A_m(\eta +\sin \eta).
\end{eqnarray}
The surface of the star is at some radial coordinate $\chi=\chi_s$ and
the parameter $\eta$ varies between $0$ and $\pi$.  The exterior
metric is given by the Schwarzschild line element,
\begin{equation} \label{ext_metric}
ds^2 = -\left(1-{2M\over{r_s}}\right)\,dt^2 + \left(1 - {2M\over
{r_s}}\right)^{-1}\,dr_s^2 + r_s^2d\Omega^2,
\end{equation}
where $M$ is the star's total mass.  The surface of the star is at
(areal) radius $r_{s}=R_s(\tau)$ and follows a radial geodesic according to
\begin{eqnarray} 
R_s & = & {1\over 2}R_s(0)(1+\cos\eta), \\ 
\tau & = & \left({R_s^3(0)\over {8M}}\right)^{1/2}(\eta +\sin\eta).
\end{eqnarray}
Matching the interior and exterior solutions at the surface yields 
\begin{eqnarray}
A_m & = & \left({R_s^3(0)\over{2M}}\right)^{1/2} \\
\sin\chi_s & = & \left({2M \over R_s(0)}\right)^{1/2} \label{chi_s}
\end{eqnarray}
According to the above equation, $\chi_s$ must lie in the range 
$0 \leq \chi_s \leq \pi/2$. 
 
The fluid 4-velocity $u^a$ satisfies the
geodesic equations,
\begin{equation} 
u^b \nabla_b u^a = 0.
\end{equation}
The rest mass-energy density $\rho$ measured in the comoving frame,
remains homogeneous and is given by
\begin{equation} \label{rho_rel}
\frac{\rho}{\rho(0)} = Q^{-3}(\tau),
\end{equation}
where 
\begin{eqnarray} 
Q(\tau) & = & {A\over {A_m}} = {1\over 2}(1 +\cos\eta)  \label{Q}\\
  \tau &= & \left({R_s^3(0)\over{8M}}\right)^{1/2}(\eta + \sin\eta) 
\label{int_tau}
\end{eqnarray}
for $ 0 \leq \eta \leq \pi$.  The proper time to collapse to a
singularity at the origin is $\tau_{coll}=\pi(R_s^3(0)/8M)^{1/2}$.
 
\section{The Interior Electrodynamic Fields}
\label{sec4}

\subsection{The Newtonian Solution}
\label{sec4.1}

The Newtonian solution describing spherical, homogeneous dust 
collapse has the same functional form as the relativistic solution
presented in Section \ref{sec3}.  In particular, the density 
satisfies
\begin{equation} \label{rho_newton}
\frac{\rho}{\rho(0)} = Q^{-3}(t),
\end{equation}
where $Q$ as a function of Newtonian time $t$ is given by (\ref{Q}) and
(\ref{int_tau}) if $\tau$ is replaced by $t$, and the velocity is
\begin{equation}
{\bf v} = \frac{\dot Q(t)}{Q(t)} {\bf r}.
\end{equation}
Here the dot denotes a derivative with respect to time.

We assume the interior of the collapsing dust ball to be
perfectly conducting, in which case the electric field vanishes
in the comoving frame,
\begin{equation}
{\bf E} = 0,
\end{equation}
and the magnetic field can be treated in the ideal
magnetohydrodynamics approximation.  It then satisfies the 
induction equation
\begin{equation} \label{induction_newton}
\partial_t {\bf B} = {\bf \nabla \times } ( {\bf v \times B} )
\end{equation}
and the divergence (or constraint) equation
\begin{equation} \label{constraint_newton}
{\bf \nabla \cdot B } = 0.
\end{equation}
In index notation in a coordinate basis, equation (\ref{induction_newton}) 
may be written
\begin{eqnarray}
\partial_t B^i & = & D_j ( v^i B^j - v^j B^i) \nonumber \\
 	& = & \gamma^{-1/2} \partial_j 
	\left( \gamma^{1/2} (v^i B^j - v^j B^i) \right),
\end{eqnarray}
where $D_i$ is the covariant derivative associated with the spatial
metric $\gamma_{ij}$, and where $\gamma$ is its determinant.  
In flat spacetime with $\partial_t \gamma = 0$, the last equation can
also be expressed as
\begin{equation} \label{ind2_newton}
\partial_t {\mathcal B}^i =  \partial_j 
	(v^i {\mathcal B}^j - v^j {\mathcal B}^i),
\end{equation}
where ${\mathcal B}^i \equiv \gamma^{1/2} B^i$ is the magnetic vector
density.  This expression is identical to the relativistic result
(4.13) of Paper I, which hereafter we will refer to as (I.4.13).  In
terms of ${\mathcal B}^i$ the constraint equation
(\ref{constraint_newton}) can be written
\begin{equation}
\partial_i {\mathcal B}^i = 0,
\end{equation}
which is the same as the relativistic expression (I.4.14).

In flat-space Cartesian coordinates we have
\begin{equation}
v^i = \frac{\dot Q}{Q} x^i
\end{equation}
and $\gamma =1$, hence ${\mathcal B}^i = B^i$.  Equation
(\ref{ind2_newton}) can then be expanded as
\begin{eqnarray}
\partial_t { B}^i & = & 
	v^i_{~,j} { B}^j 
	+ v^i { B}^j_{~,j} 
	- v^j_{~,j} { B}^i
	- v^j { B}^i_{~,j} \nonumber \\
& = &	\frac{\dot Q}{Q} { B}^i 
	- 3 \frac{\dot Q}{Q} { B}^i 
	- v^j { B}^i_{~,j}.
\end{eqnarray}
Since the total derivative following fluid elements is
\begin{equation}
\frac{d}{dt} = \frac{\partial}{\partial t} 
	+ v^i \frac{\partial}{\partial x^i}
\end{equation}
we find
\begin{equation}
\frac{d}{dt}  { B}^i = - 2 \frac{\dot Q}{Q} { B}^i
\end{equation}
with implies
\begin{equation}
\frac{B^{i}(t)}{B^{i}(0)} = Q(t)^{-2} =
\left(\frac{\rho}{\rho(0)}\right)^{2/3} = 
\left(\frac{R_s(0)}{R_s}\right)^2.
\end{equation}
Here we have used (\ref{rho_newton}), and 
$R_{s}$ and
$R_s(0)$ are the radius of the star and its initial value,
respectively.

\subsection{The Relativistic Solution}
\label{sec4.2}

The relativistic interior solution is most easily evaluated in the
comoving coordinates of the metric (\ref{int_metric}).  In this
coordinate system the fluid four-velocity is
\begin{equation}
u^a = \left( \frac{\partial}{\partial \tau} \right)^a
\end{equation}
or 
\begin{equation}
u^0 = 1,~~~~~u^i = 0.
\end{equation}
In these comoving coordinates the relativistic induction equation
(I.4.13), which is identical to the Newtonian version
(\ref{ind2_newton}), reduces to
\begin{equation} \label{B_int_1}
\frac{\partial}{\partial \tau} {\mathcal B}^i = 
\frac{\partial}{\partial \tau} (\gamma^{1/2} B^i) = 0.
\end{equation}
For the metric (\ref{int_metric}) we have
\begin{equation} \label{gamma_1}
\gamma^{1/2} = A^3 \sin^2 \chi \sin \theta
\end{equation}
where $A = A(\tau)$ is given by equations~(\ref{Q}) and
(\ref{int_tau}). The components of $B^i$ are most easily expressed in
an orthonormal basis.  The orthonormal basis vectors ${\bf e}_{\hat
i}$ can be written in terms of the coordinate basis vectors as
\begin{eqnarray}
{\bf e}_{\hat \chi} & = & \frac{1}{A} {\bf e}_{\chi} 
	= \frac{1}{A} \frac{\partial}{\partial \chi} \nonumber \\
{\bf e}_{\hat \theta} & = & \frac{1}{A \sin \chi} {\bf e}_{\theta} 
	= \frac{1}{A \sin \chi} \frac{\partial}{\partial \theta} \\
{\bf e}_{\hat \phi} & = & \frac{1}{A \sin \chi \sin \theta} {\bf e}_{\phi} 
	= \frac{1}{A\sin \chi \sin \theta} \frac{\partial}{\partial \phi} 
\nonumber
\end{eqnarray}
As a consequence, all orthonormal components $B^{\hat i}$ are
proportional to $A B^i$.  Inserting these together with
(\ref{gamma_1}) into (\ref{B_int_1}) then yields
\begin{equation}
B^{\hat i} A^2 = \mbox{const}
\end{equation}
or
\begin{equation} \label{Bintsol}
\frac{B^{\hat i}(\tau)}{B^{\hat i}(0)} = Q^{-2}(\tau) =
\left(\frac{\rho}{\rho(0)}\right)^{2/3} = 
\left(\frac{R_s(0)}{R_s}\right)^2
\end{equation}
where we have used (\ref{rho_rel}).  The above equation holds for 
{\it any} initial field configuration $B^{\hat i}(0)$.  As in the Newtonian 
case, the electrical field $E^i$ vanishes in the comoving frame 
under the assumption of perfect conductivity,
\begin{equation} \label{Emhd}
E^{\hat i}(\tau) = 0.
\end{equation}

\section{The Exterior Electrodynamic Fields}
\label{sec5}

\subsection{Maxwell's equations}
\label{sec5.1}

The exterior of the collapsing dust sphere is pure vacuum, so that the
approximation of ideal magnetohydrodynamics cannot be applied there.
Instead, Maxwell's equations have to be solved without approximation.
In terms of the electric field $E_i$ and the vector potential $A_i$ as
measured by a normal observer $n^a$, Maxwell's equations can be
written
\begin{equation} \label{Adot1}
\partial_t A_i = - \alpha E_i + \beta^j A_{i,j}
	+ \beta^j_{~,i} A_j - \partial_i (\alpha \Phi)
\end{equation}
(equation (I.3.22)) and 
\begin{eqnarray} \label{Edot1}
\partial_t E^i & = & \gamma^{-1/2} \left( \alpha \gamma^{1/2} (\gamma^{il}
\gamma^{jm} - \gamma^{im} \gamma^{jl}) A_{m,l} \right),_j \nonumber \\
& & - 4 \pi \alpha J^i + \alpha K E^i + \beta^j E^i_{~,j} 
	- E^j \beta^i_{~,j}
\end{eqnarray}
(equation (I.3.25)).  Without loss of generality, we choose the
source-free electromagnetic Coulomb gauge condition
\begin{equation} \label{gauge}
\Phi = 0.
\end{equation}

In Section \ref{sec7} we will present numerical solutions to equations
(\ref{Adot1}) and (\ref{Edot1}) in three different time slicings and
spatial coordinate systems: Schwarzschild time slicing in isotropic
coordinates, maximal slicing in isotropic coordinates, and Kerr-Schild
time slicing in Kerr-Schild spatial coordinates.  For the first two
spacetime metrics, the metric takes the spherically symmetric
isotropic form
\begin{equation} \label{metric_isotropic}
ds^2 = - (\alpha^2 - A^2 \beta^2) dt^2 + 2 A^2 \beta dr dt
	+ A^2 (dr^2 + r^2 d \Omega^2),
\end{equation}
where $\beta = \beta^r$, $\alpha$ and $A$ are all functions
of $r$ and $t$ alone, and $K=0$. We will focus on this form here, and
will generalize to a nonisotropic Kerr-Schild metric in Section
\ref{sec7.4}.

In this paper we shall consider axisymmetric (dipole) field
configurations.  Complete axisymmetric solutions exist for which the
only non-vanishing components of $A_i$ and $E^i$ are $A_{\phi} =
A_{\phi}(t,r,\theta)$ and $E^{\phi} = E^{\phi}(t,r,\theta)$.
(Correspondingly, the only non-vanishing components of $B^i$ are
$B^r(t,r,\theta)$ and $B^{\theta}(t,r,\theta)$). Inserting these
quantities together with (\ref{gauge}) and (\ref{metric_isotropic})
into equations (\ref{Adot1}) and (\ref{Edot1}) yields
\begin{equation} \label{Adot2}
\partial_t A_{\phi} = - \alpha E_{\phi} + \beta \partial_r A_{\phi}
\end{equation}
and
\begin{eqnarray} \label{Edot2}
\partial_t E^{\phi} & = &
	- \frac{1}{A^3 r^2 \sin \theta} \left(
	\left( \frac{\alpha}{A} A_{\phi,r} \right)_{,r} 
	+ \left( \frac{\alpha}{A r^2 \sin \theta} A_{\phi,\theta} 
	\right)_{,\theta} \right) \nonumber \\ 
& & 	+ \beta \partial_r E^{\phi}.
\end{eqnarray}

The fields $A_{\phi}$ and $E^{\phi}$ can now be decomposed into
multipoles.  A particularly convenient decomposition is
\begin{equation} \label{A_expansion}
A_{\phi}(t,r,\theta) = - \sum_l a^l(t,r) 
\left( (1 - \mu^2) \frac{dP_l(\mu)}{d\mu} \right),
\end{equation}
where $\mu = \cos \theta$ and where the $P_l(\mu)$ are the Legendre
polynomials of order $l$.  Inserting this into (\ref{Edot2}) yields
\begin{eqnarray} \label{Edot3}
\partial_t E^{\phi} & = & 
\frac{1}{A^3 r^2} \left( \frac{dP_l(\mu)}{d\mu} \right)
	\left( \frac{\alpha}{A} a^l_{~,r} \right)_{,r} \\
& & +  \frac{\alpha}{A^4 r^4} a^l \left( (1 - \mu^2) \frac{dP_l(\mu)}{d\mu}
	\right)_{,\mu\mu} + \beta \partial_r E^{\phi},\nonumber
\end{eqnarray}
where a summation over repeated $l$ is implied. 
Since the $P_l(\mu)$ satisfy
\begin{equation}
\frac{d}{d\mu} \left( (1 - \mu^2) \frac{d P_l}{d\mu} \right)
= - l (l + 1) P_l,
\end{equation}
equation (\ref{Edot3}) can rewritten as
\begin{equation} \label{Edot4}
\partial_t E^{\phi} =
\left(\frac{1}{A^3 r^2} \left( \frac{\alpha}{A} a^l_{~,r} \right)_{,r}
	- \frac{l (l+1)}{r^4} \frac{\alpha}{A^4} a^l \right) \frac{dP_l}{d\mu}
	+ \beta \partial_r E^{\phi}.
\end{equation}

We now decompose $E^{\phi}$ as
\begin{equation} \label{E_expansion}
E^{\phi}(t,r,\theta) = \sum_l \frac{e^l(t,r)}{r^2} \frac{dP_l}{d\mu}.
\end{equation}
The covariant component $E_{\phi}$ is then
\begin{equation} 
E_{\phi} = \sum_l e^l(t,r) A^2 (1 - \mu^2) \frac{dP_l}{d\mu}.
\end{equation}
This expression can be inserted into equations (\ref{Adot2}) and
(\ref{Edot4}), which, after dropping the superscript $l$ in $a^l$ and
$e^l$, yields the coupled set of differential equations in $t$ and
$r$ for each mode $l$:
\begin{eqnarray}
\partial_t e & = & \frac{1}{A^3} \left( \frac{\alpha}{A} a_{,r} \right)_{,r}
	- \frac{\alpha}{A^4} \frac{l (l+1)}{r^2} \, a \nonumber \\
   & & + \beta r^2 \left( \frac{e}{r^2} \right)_{,r} \label{edot1} \\
\partial_t a & = & \alpha A^2 e + \beta a_{,r}. \label{adot1}
\end{eqnarray}
In the following we will specialize to pure dipole fields with $l=1$.

\subsection{Boundary Conditions}
\label{sec5.2}

The dynamical fields $a$ and $e$ satisfy boundary conditions both on
the surface of the star and on the outer edge of the numerical grid,
which we take to be at a large separation from the star.  On the
surface of the star, the fields have to be matched to the interior
solution (Section \ref{sec5.2.1}), except when the stellar interior
collapses inside the event horizon and is excised from the numerical
grid (Section \ref{sec7.4}). At large distance we impose ``outgoing
wave boundary conditions'' (Section \ref{sec5.2.2}).

\subsubsection{Inner Boundary Conditions}
\label{sec5.2.1}

In a frame comoving with the surface of the star, the component of the
magnetic field normal on the surface and the tangential components of
the electric field have to be continuous across this surface.  To
derive boundary conditions for $a$ and $e$, we therefore first have to
boost from our computational frame of normal observers $n^a$ to a
frame comoving with the surface of the star $u^a$. We then have to
express the transformed junction conditions on the orthonormal
components of the magnetic and electric fields in terms of the
expansion coefficients $a$ and $e$.

Denote the normal observer frame by $F'$ and the frame comoving with
the stellar surface by $F$. Boosting with velocity $v$ from $F$ to
$F'$ leads to the following relation between the orthonormal
components of the magnetic and electric fields in the two frames:
\begin{equation}
\begin{array}{rclcrcl} \label{EBtrans}
E_{\hat r'} & = & E_{\hat r} & ~~ &
B_{\hat r'} & = & B_{\hat r} \\
E_{\hat \theta'} & = & \gamma (E_{\hat \theta} - v B_{\hat \phi}) & & 
B_{\hat \theta'} & = & \gamma (B_{\hat \theta} + v E_{\hat \phi}) \\
E_{\hat \phi'} & = & \gamma (E_{\hat \phi} + v B_{\hat \theta}) & & 
B_{\hat \phi'} & = & \gamma (B_{\hat \phi} - v E_{\hat \theta}) .\\
\end{array}
\end{equation}
Here $\gamma$ is the gamma-factor between the two 
frames $F'$ and $F$,
\begin{equation}
\gamma = - u_a n^a = (1 - v^2)^{-1/2}.
\end{equation}
In the comoving frame $F$, the relevant junction conditions for
$B_{\hat r}$ and $E_{\hat \phi}$ at the surface 
are (see, e.g., Eqs.~(21.161a-d) in Misner, Thorne \& Wheeler 1973)
\begin{eqnarray}
B_{\hat r}^{(in)} & = & B_{\hat r}^{(out)} \label{B_cond_in} \\
E_{\hat \phi}^{(in)} & = & E_{\hat \phi}^{(out)} = 0, \label{E_cond_in}
\end{eqnarray}
where the electric field component has to vanish because of the assumption
of MHD in the interior, equation~(\ref{Emhd}).  Combining the Lorentz 
transformations (\ref{EBtrans}) with equation~(\ref{B_cond_in}) 
yields the following condition for the magnetic field 
in the frame $F'$ of a normal observer:
\begin{equation} \label{Bbc}
B_{\hat r'}^{(out)} = B_{\hat r}^{(in)}.
\end{equation}
To derive a condition on the electric field $E_{\hat \phi'}^{(out)}$,
it is useful to consider the inverse transformation
\begin{equation} \label{inv}
E_{\hat \phi}^{(out)} = \gamma (E_{\hat \phi'}^{(out)}
	- v B_{\hat \theta'}^{(out)}).
\end{equation}
According to (\ref{E_cond_in}) the left hand side of (\ref{inv}) 
has to vanish, which yields the following boundary condition on the
electric field in $F'$:
\begin{equation} \label{Ebc}
E_{\hat \phi'}^{(out)} = v B_{\hat \theta'}^{(out)}.
\end{equation}

We now express the orthonormal components of the exterior magnetic and
electric field in terms of the quantities $a$ and $e$.  The magnetic
field $B^i$ can be found from
\begin{equation}
B^i = \epsilon^{ijk} A_{k,j}
\end{equation}
(see, e.g., Eq.~(I.3.17)).  The Levi-Civita tensor can be expressed
as $\epsilon^{ijk} = \gamma^{-1/2} [ijk]$, where $[ijk]$ is the 
completely antisymmetric symbol, from which we find
\begin{equation}
B^r = \frac{1}{A^3 r^2 \sin \theta} A_{\phi,\theta}
\end{equation}
and
\begin{equation}
B^{\theta} = - \frac{1}{A^3 r^2 \sin \theta} A_{\phi,r}
\end{equation}
for the metric (\ref{metric_isotropic}).  Specializing to pure dipole
fields with $l = 1$, the expansion (\ref{A_expansion}) for $A_{\phi}$
reduces to
\begin{equation} \label{mom}
A_{\phi} = - a \sin^2 \theta,
\end{equation}
and we therefore have
\begin{equation} \label{Br}
B^{\hat r} = A B^r = - \frac{2 \cos \theta}{A^2 r^2} \, a
\end{equation}
and
\begin{equation} \label{Btheta}
B^{\hat \theta} = A r B^{\theta} = \frac{\sin \theta}{A^2 r} \, a_{,r}.
\end{equation}
For the exterior electric field $E^{\hat \phi}$ we similarly find
\begin{equation} \label{Ephi}
E^{\hat \phi} = A r \sin \theta E^{\phi} = \frac{A \sin \theta}{r} \, e.
\end{equation}

Inserting the last three expressions into the boundary
conditions~(\ref{Bbc}) and~(\ref{Ebc}) on the fields, we finally
obtain
\begin{equation} \label{a_inner_condition}
a = - A^2 r^2 \frac{B_{\hat r}^{(in)}}{2 \cos \theta}
\end{equation}
and
\begin{equation} \label{e_inner_condition}
e = \frac{v}{A^3} a_{,r}.
\end{equation}

For simulations in Kerr-Schild coordinates, we excise the interior
of the black holes at late times.  In this case, these inner boundary
conditions have to be replaced with boundary conditions on the surface
of the excised region, as we will discuss in Section \ref{sec7.4}.

\subsubsection{Outer Boundary Conditions}
\label{sec5.2.2}

In the asymptotically flat region at large distances from the star,
equations~(\ref{edot1}) and~(\ref{adot1}) combine to give
\begin{equation} \label{asym}
a_{,tt} - a_{,rr} - \frac{l(l+1)a}{r^2} = 0,
\end{equation}
and a similar equation for $e$. For our adopted $l=1$ initial exterior
magnetic dipole field (see Section \ref{sec6.2}), the solution to
equation~(\ref{asym}) is characterized by a longitudinal dipole field
$a \sim 1/r$ at early times and an outgoing radiation field $a \sim
f(t-r)$ at late times. A suitable but approximate boundary condition
which we adopt to reflect this behavior in the wave zone $r \gg
\lambda \sim 25 M$ is of the form
\begin{equation} \label{outer_boundary}
e,a \sim \frac{f(t-r)}{r},
\end{equation}
Finite difference implementations of this equation can be found in
Appendix \ref{secB}.  In fact, we find that most of our results are
quite insensitive to the precise form of our outer boundary
conditions. The reason is that the longitudinal dipole field has a
vanishingly small value at our outer boundary initially and our
integrations terminate before any electromagnetic wave ever reaches
the outer boundary.

\section{Electrodynamic Initial Data}
\label{sec6}

As initial data for the electromagnetic fields we seek the
relativistic generalization of the Newtonian data describing a
uniformly magnetized sphere of radius $R$, matched onto an exterior
dipole magnetic field.  In orthonormal polar coordinates, the interior
solution is given by the constant field
\begin{equation} \label{BinitN}
B^{\hat i} = B (\cos \theta, -\sin \theta, 0)~~~~~~\mbox{(Newtonian
interior)}
\end{equation}
for the field aligned with the $z$-axis, and the exterior solution for
a pure dipole field is given by 
\begin{equation} \label{BinitNext}
B^{\hat i} = \frac{BR^3}{r^3} (\cos \theta, \frac{1}{2} \sin \theta,
0)~~~~~~\mbox{(Newtonian exterior)}.
\end{equation}
The junction conditions on the surface of the star require that the
normal (i.e.~radial) component of the field be continuous across the
stellar surface.  The tangential discontinuity in $B^i$ implies the
presence of a surface current.

\subsection{Relativistic Interior Field}
\label{sec6.1}

The relativistic generalization of the Newtonian interior solution
(\ref{BinitN}) has to satisfy the constraint equation (I.3.9)
\begin{equation}
D_i B^i = 0
\end{equation}
where $D_i$ is the covariant derivative associated with the interior
metric (\ref{int_metric}). A suitable solution is 
\begin{eqnarray}
B^{\hat \chi} & = & B \cos \theta 
	\left( \frac{\chi}{\sin \chi} \right)^2
	\left( \frac{\chi_s}{\sin \chi_s} \right)^{-2} \label{Bchiint} \\
B^{\hat \theta} & = & - B \sin \theta 
	\left( \frac{\chi}{\sin \chi} \right)
	\left( \frac{\chi_s}{\sin \chi_s} \right)^{-2}, \label{Bthetaint}
\end{eqnarray}
where $B$ is an arbitrary constant in space. This solution
automatically satisfies the magnetic induction equation during the
stellar collapse if $B$ is chosen to vary with proper time according
to
\begin{equation} \label{Bbc2}
B = B_0 \left( \frac{\rho}{\rho(0)} \right)^{2/3}
\end{equation}
(see Section \ref{sec4.2} and equation (\ref{Bintsol})).  

At the surface $\chi = \chi_s$, the radial component $B^{\hat \chi}$,
which can be identified with $B^{\hat r}$, reduces to
\begin{equation} \label{Bchiintsurf}
B^{\hat \chi} = B \cos \theta~~~~~~\mbox{(at surface)},
\end{equation}
which we will use in the boundary condition (\ref{a_inner_condition}).

Finally, it is easy to see that the components (\ref{Bchiint}) and
(\ref{Bthetaint}) reduce to the Newtonian solution (\ref{BinitN}) in
the Newtonian limit $2M/R_s \ll 1$. From (\ref{chi_s}), this inequality 
implies that $\sin \chi_s \ll 1$ and hence $\chi_s \approx \sin \chi_s$. If
this holds for the surface value $\chi_s$, it must also hold for all
interior values $\chi \leq \chi_s$. All the fractions in equations
(\ref{Bchiint}) and (\ref{Bthetaint}) therefore reduce to unity, so
that we recover the Newtonian solution (\ref{BinitN}).

\subsection{Relativistic Exterior Field}
\label{sec6.2}

An analytic solution for the magnetic field in the exterior of a
static spherical star as measured by a static observer has been given
by Wasserman \& Shapiro (1983):
\begin{eqnarray}
B_{\hat r} & = & - \frac{6 \mu_d \cos \theta}{r_{s}^3}
	x_{s}^2 \Big( x_{s} \ln(1 - x_{s}^{-1}) \nonumber \\
& & + (1 + x_{s}^{-1}/2) \Big) \label{Brext} \\
B_{\hat \theta} & = & \frac{6 \mu_d \cos \theta}{r_{s}^3} x_{s}^2 
	\Big( x_{s} (1 - x_s^{-1})^{1/2} \ln(1 - x_{s}^{-1}) \nonumber \\
& & + \frac{1 - x_s^{-1}/2}{(1 - x_s^{-1})^{1/2}} \Big) 
\label{Bthetaext}
\end{eqnarray}
Here $r_{s}$ is the Schwarzschild (areal) radius, $x_{s} \equiv
r_{s}/(2 M)$, and $\mu_d$ is the magnetic dipole moment.  Using the
junction condition (\ref{B_cond_in}), $\mu_d$ can be expressed in
terms of the interior field by matching the interior and exterior
components of $B_{\hat r}$ (equations (\ref{Bchiint}) and
(\ref{Brext})) on the surface.  This yields
\begin{equation} \label{mu_d}
\mu_d = - B \frac{R_{s}^3}{6 X_{s}^2} 
	\left(X_{s} \ln(1 - X_{s}^{-1}) + (1 - X_{s}^{-1}/2) \right)^{-1},
\end{equation}
where $R_{s}$ and $X_{s}$ are the values of $r_s$ and $x_s$ on the
stellar surface, and where $B$ is the factor appearing in equation
(\ref{Bchiint}).  In the Newtonian limit $x_s \gg 1$ the solutions
(\ref{Brext}) and (\ref{Bthetaext}) with $\mu_d$ given by (\ref{mu_d})
reduces to the Newtonian solution (\ref{BinitNext}).

From equations (\ref{Brext}) and (\ref{Bthetaext}) the magnetic
potential $A_{\phi}$ can be found to be
\begin{equation}
A_{\phi} = - \frac{3 \mu_d \sin^2 \theta}{2M} 
	\left(x_s^2 \ln(1 - x_s^{-1}) + x_s + \frac{1}{2} \right),
\end{equation}
and therefore, according to equation (\ref{mom}), 
\begin{equation}
a = \frac{3 \mu_d}{2M} 
	\left(x_s^2 \ln(1 - x_s^{-1}) + x_s + \frac{1}{2} \right).
\end{equation}

We take the above form for the exterior magnetic field as initial data
in our numerical simulations. We also set
\begin{equation}
E^{\hat i} = 0 = e, 
\end{equation}
which is consistent with our star having zero charge.  For
Schwarzschild slicing and maximal slicing the normal observer is a a
static observer at $t=0$, a moment of time symmetry, so the components
of $B^{\hat i}$ and $E^{\hat i}$ as measured by $n^a$ are identical
to the values quoted above at the initial time (see Section
\ref{sec7.4.3} for the case of Kerr-Schild slicing).  However, since
our dynamical Oppenheimer-Snyder spacetime is not the spacetime of a
static star, the above initial data no longer yields a static magnetic
field solution. However, it does provide a physically plausible field
for our (arbitrary) initial data, if we assume that star was dominated
by a dipole field up to the moment of collapse.

\section{Exterior Electromagnetic Field Solution in Different Time Slicings}
\label{sec7}

We now integrate the coupled equations~(\ref{edot1}) and~(\ref{adot1})
for the exterior electric field $E^{\phi}$ and the magnetic potential
$A_{\phi}$, given the interior solution and initial data as in
Sections \ref{sec4} and \ref{sec6}.

This problem provides an ideal laboratory for experimenting with
various coordinate gauge choices for performing numerical integrations in strong
gravitational fields.  We will compare three different time slicings
for treating the exterior evolution, namely Schwarzschild, maximal and
Kerr-Schild. Historically, the earliest numerical studies
traditionally worked in the Schwarzschild gauge when dealing with
spherical, asymptotically flat spacetimes. But in this gauge the time
coordinate approaches infinity as the stellar surface reaches the
black hole event horizon at areal radius $r_s = 2M$; light cones pinch
off near the horizon and the resulting coordinate singularities make
integration of the electromagnetic field equations to arbitrarily late
times difficult in this gauge.  Maximal time slicing has often been
adopted in recent numerical work as a ``singularity avoiding'' gauge;
spatial slices cover the entire matter interior as well as the vacuum
exterior and penetrate the horizon.  But ``grid stretching'' along the
black hole throat (see, e.g., Shapiro \& Teukolsky 1986) as the
stellar surface approaches a limiting surface inside the horizon makes
the solution increasingly inaccurate at late times.  Maximal time
slicing requires the numerical solution of a second order elliptic
equation for the lapse. While boundary conditions on this equation are
straightforward to impose at the stellar origin and at large distances
from the matter field, they are not at all simple to impose at a black
hole horizon. Consequently, it is not straightforward in this gauge to
``excise'' the black hole interior from the computational grid, even
though this region (where inaccuracies are growing) is causally
disconnected from the exterior.  These problems can be avoided in
Kerr-Schild slicing, where the lapse is nonzero on the horizon (unlike
Schwarzschild but like maximal) and analytic (like Schwarzschild but
unlike maximal). These features allow for horizon penetration and
``black hole excision'' (Unruh, unpublished; Thornburg 1987; Seidel \&
Suen 1992) to follow the evolution of the electromagnetic fields
outside the black hole to arbitrary late times.

\subsection{Analytic Precursor: Quasi-Static Approach}
\label{sec7.1}

In the earliest attempts to calculate the exterior magnetic field of a
magnetized star undergoing stellar collapse it was assumed that the
exterior field is a (longitudinal) dipole field which changes {\it
quasi-statically} during the entire collapse.  In this approximation,
the electric field and the time derivative of the magnetic field is
neglected, and the exterior magnetic field is given, at all times, by
equations (\ref{Brext}) -- (\ref{Bthetaext}).  The magnetic field
strength is determined by the magnetic dipole moment $\mu_d$, which
can be found from equation (\ref{mu_d}).  As the stellar surface $R_s$
approaches the horizon, the magnetic dipole moment approaches zero
according to
\begin{equation}
\mu \propto \frac{1}{\ln(1 - 2M/R_s)} \mbox{~~~as~~~} R_s \rightarrow 2M.
\end{equation}
During this late phase of the collapse we have
\begin{equation}
1 - \frac{2M}{R_s} \propto \exp(-t/4M)  \mbox{~~~as~~~} R_s \rightarrow 2M
\end{equation}
(see, e.g.~Problem 15.11 in Lightman et al. 1975).
Together, these imply
\begin{equation}
\mu \propto t^{-1} \mbox{~~~as~~~} R_s \rightarrow 2M,
\end{equation}
suggesting that the exterior magnetic field should decay with $t^{-1}$
at late times (see, e.g., Ginzburg \& Ozernoy 1965; Anderson \&
Cohen 1970; Zeldovich \& Novikov 1971).  This result is incorrect,
since the time-changing magnetic dipole in the collapsing star
generates a transverse electromagnetic field ($E$ and $B$), which
dominates the late-time behavior but is ignored in this quasi-static
approximation.

The correct decay rate at late times of an initially static dipole
electromagnetic radiation field outside a black hole is actually
$t^{-(2l+2)}$ (Price 1972a,b; also pointed out by Thorne 1971).  Our
numerical integrations below confirm this result for $l=1$. In
addition, we are able to obtain the complete interior and exterior
numerical solution over the entire collapse of a magnetic star to a
black hole and show the transition from a quasi-static longitudinal to
a dynamic transverse radiation field.

\subsection{Schwarzschild Slicing}
\label{sec7.2}

\begin{figure*}[t]
\begin{center}
\leavevmode
\epsfxsize=5.5in
\epsffile{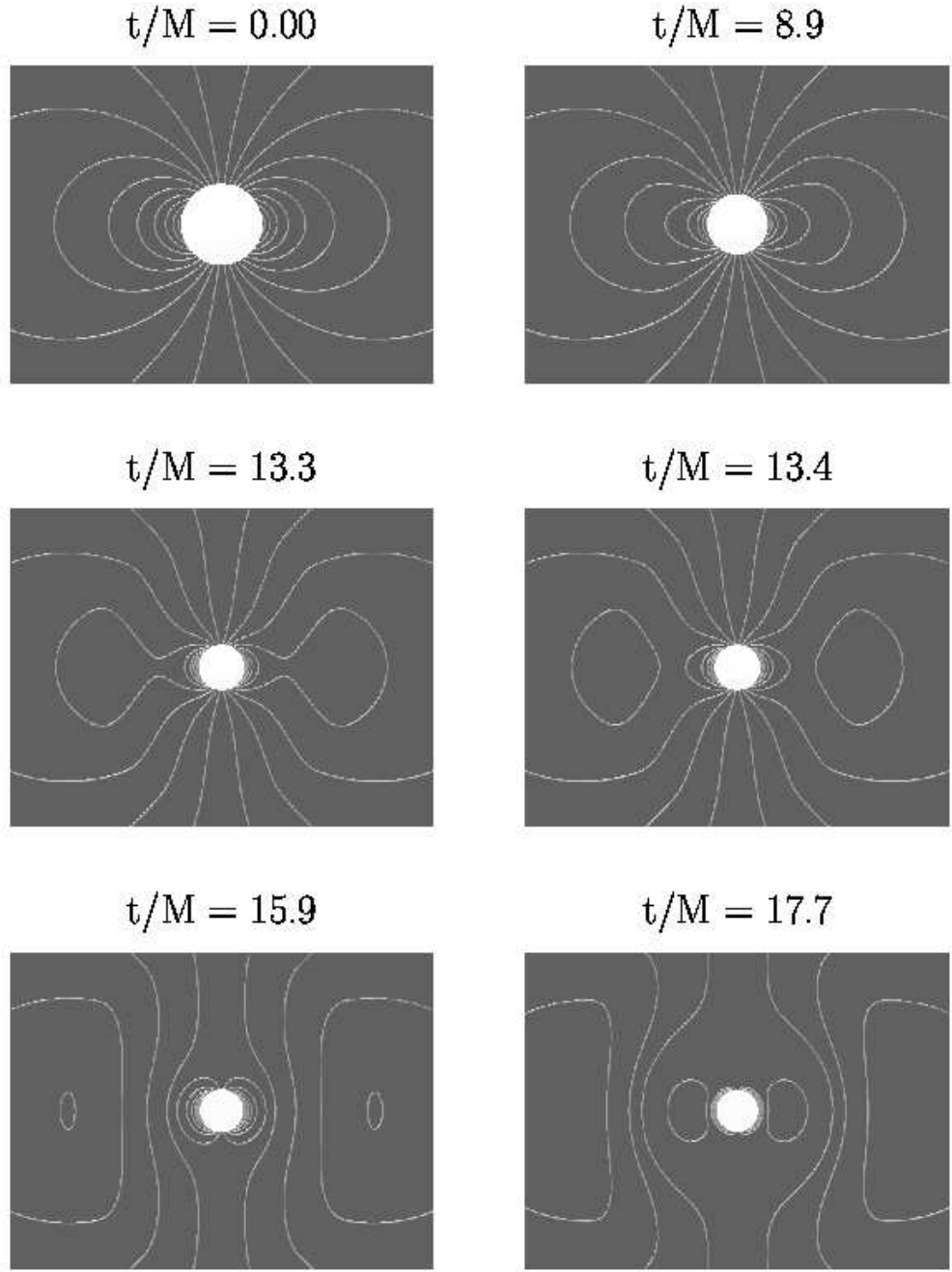}
\end{center}
\caption{Snapshots of the exterior dipole magnetic field lines at
select {\it Schwarzschild} time slices for a star which collapses from
rest at an initial areal radius $R_s = 4 M$. Points are plotted in a
meridional plane using areal radii. The light shaded sphere covers the
matter interior.  The initial growth of the longitudinal field due to
flux freezing in the interior is ultimately followed by a outward
burst of transverse electromagnetic radiation as the star approaches
the black horizon at $r_s = 2M$.  Soon after the last snapshot at $t =
17.7 M$ the quality of the numerical integration deteriorates in this
gauge.  See text for details.}
\label{fig1}
\end{figure*}

\begin{figure*}[t]
\begin{center}
\leavevmode
\epsfxsize=5.5in
\epsffile{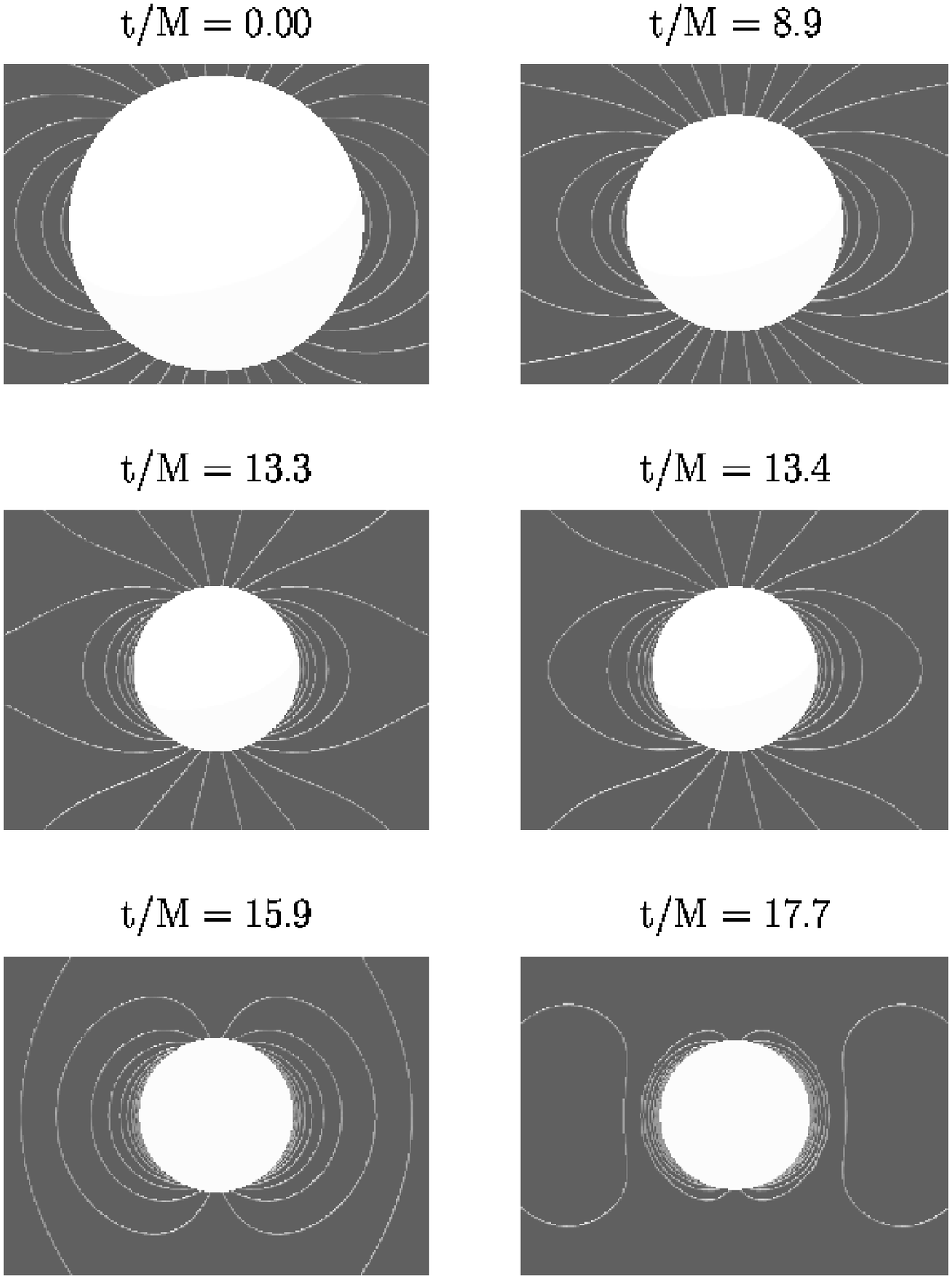}
\end{center}
\caption{Blow-up of the same collapse depicted in Figure \ref{fig1}. Note 
how in this gauge the exterior field lines which are linked to the 
interior are crushed tangentially toward the surface as it approaches 
the horizon. Field lines which pinch off and become disjoint from the 
interior propagate outward as a dipole electromagnetic wave.}
\label{fig2}
\end{figure*}

In Schwarzschild time slicing and isotropic coordinates, the exterior 
metric takes the 
form (\ref{metric_isotropic}) with the lapse $\alpha$ given by
\begin{equation} \label{alpha_SS}
\alpha = \frac{1 - M/(2r)}{1 + M/(2r)},
\end{equation}
the conformal factor $A$ given by
\begin{equation} \label{A_SS}
A = \left(1 + \frac{M}{2r} \right)^2,
\end{equation}
and with the shift $\beta$ vanishing identically,
\begin{equation} \label{beta_SS}
\beta = 0.
\end{equation}
In the above, the isotropic radius $r$ is related to the areal radius
$r_s$ by
\begin{equation}
r_s = r \left(1 + \frac{M}{2r} \right)^2
\end{equation}
and 
\begin{equation} \label{r(r_a)}
r = \frac{1}{2} \left(r_s - M + \left(r_s (r_s - 2 M) \right)^{1/2} \right).
\end{equation}
The above
quantities appear in the evolution equations (\ref{edot1}) for $e$ 
and (\ref{adot1}) $a$. 

Analytic expressions for the areal radius $R_s$ of the stellar surface
as a function of proper time $\tau$ have been given in Section
\ref{sec3} in terms of the parameter $\eta$.  The isotropic radius can
be found by substituting the areal radius into (\ref{r(r_a)}).  The
Schwarzschild coordinate time $t$ can also be found in terms of $\eta$
according to
\begin{eqnarray}
t & = & 2M \ln \left( \frac{(R_s(0)/(2M) - 1)^{1/2} + \tan(\eta/2)}
	{(R_s(0)/(2M) - 1)^{1/2} - \tan(\eta/2)} \right) \nonumber \\
& & 	+ 2 M \left( \frac{R_s(0)}{2M} - 1 \right)^{1/2} 
	\left( \eta + \frac{R_s(0)}{4M}(\eta + \sin \eta) \right)
\end{eqnarray}
(see, e.g.~Misner, Thorne \& Wheeler 1973, eqn. 32.10b).  Note that we
have $t/M \rightarrow \infty$ as the stellar surface approaches the
event horizon, $R_s \rightarrow 2M$.

The velocity $v$ of the surface with respect to a normal observer is
\begin{equation}
v = \left(\frac{2M/R_s - 2M/R_s(0)}{1 - 2M/R_s(0)} 
	\right)^{1/2}.
\end{equation}
This quantity enters the inner boundary condition for $e$, equation
(\ref{e_inner_condition}).

We have solved equations (\ref{edot1}) and (\ref{adot1}) in
Schwarzschild slicing with an implicit finite differencing scheme (see
Appendix \ref{secB}).  We regrid after each timestep so that the
stellar surface is always located half-way between the two first
grid-points (which makes the implementation of the boundary conditions
(\ref{a_inner_condition}) and (\ref{e_inner_condition}) particularly
simple).  The advantage of implicit schemes is that no Courant
condition has to be satisfied to maintain stability of the numerical
integration.  Because of the linearity of Maxwell's equations the
initial field strength $B_0$ is arbitrary (provided it is dynamically
unimportant) and can, without loss of generality, be chosen to be
unity.

\begin{figure*}[t]
\begin{center}
\leavevmode
\epsfxsize=3in
\epsffile{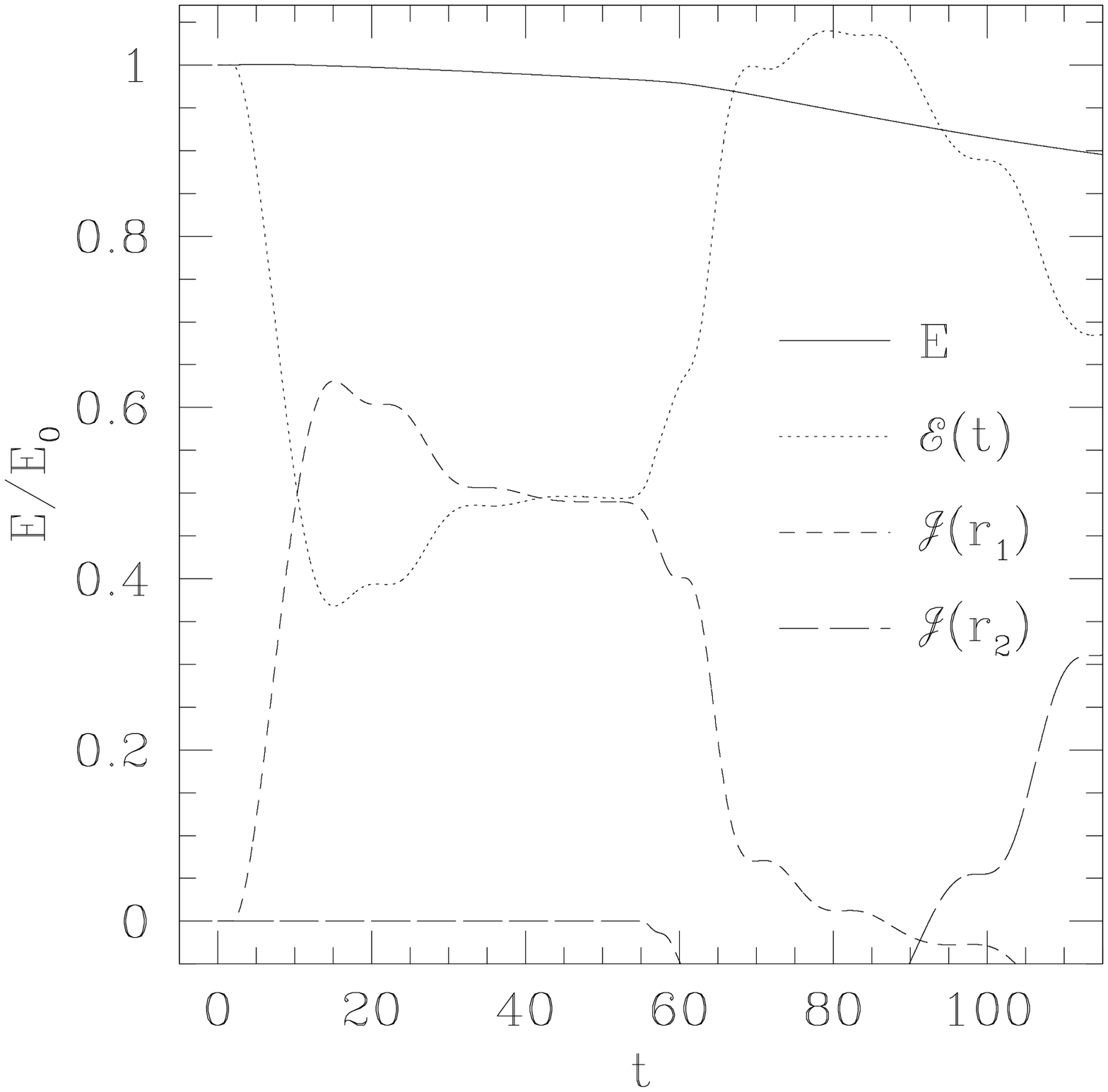}
\end{center}
\caption{Electromagnetic energy conservation in Schwarzschild time
slicing.  The outer boundary is located at $R_{\rm out} = 100 M$, the
exterior of the star is covered with 5000 gridpoints, and the Courant
factor is unity.  The solid line denotes the total energy $E$, which
should be conserved (see equation ~(\ref{Econs})).  The dotted line is
the energy ${\mathcal E}(t)$ contained between the radii $r_1 = 1.01
R_s(0) = 4.04 M$ and $r_2 = 80 M$; the short dashed line is the
integrated flux across $r_1$, ${\mathcal J}(r_1)$, and the long dashed
line is the integrated flux across $r_2$, ${\mathcal J}(r_2)$.  All
energies are normalized to the initial energy $E_0$.  Note that the
quality of the numerical integration starts to deteriorate at about $t
\sim 20 M$, when the stellar surface and the ingoing electromagnetic
waves approach the event horizon at $r_s = 2M$. }
\label{fig3}
\end{figure*}

In Figures \ref{fig1} and \ref{fig2} we show numerical results for a
star collapsing from rest from an initial radius of $R_s(0) =
4M$.  In these snapshots we show the collapsing star together with
exterior magnetic field lines.  In axisymmetry, these field lines
coincide with contours of constant $A_{\phi}$ (Zhang 1989), which
enables us to draw them easily.  At $t=0$ all field lines are attached
to fluid interior and emerge from the stellar surface.  At later
times, as the surface approaches the event horizon, field lines that
emerge from the surface are crushed against the star and become
increasingly tangential (Anderson \& Cohen 1970).  At larger
separations, field lines detach from the star during the collapse.  In
the radiation zone, these disjoint lines propagate outwards with the
speed of light.  Thus, early on, we find that the exterior
longitudinal $B$ field evolves quasi-statically to match the growth of
the frozen-in interior field (see equation~(\ref{Bintsol})), while the
exterior $E$ field remains small.  At late times, the exterior
longitudinal fields are radiated away and the dominant $B$ and $E$
fields become outgoing, decaying, transverse electromagnetic waves.

To calibrate the accuracy of the numerical integration we check the
global conservation of magnetic energy as outlined in Appendix
\ref{secA}.  We show results of this energy conservation test in
Figure \ref{fig3}.  Clearly, the accuracy deteriorates soon after the
surface of the star approaches the event horizon at around $t \sim
20M$.  This effect is not surprising.  The vanishing of the
lapse~(\ref{alpha_SS}) on the horizon causes the local light cone to
``pinch off'' and the (coordinate) speed of light to go to zero.  As a
consequence, incoming radiation piles up in front of the horizon and
develops smaller and smaller spatial structures, which ultimately
become smaller than any fixed grid resolution (cf.~Rezzolla et
at.~1998).

It is evident that Schwarzschild coordinates are not suitable for
calculating the late-time radiation as measured by a distant observer,
and, hence, for determining the late-time electromagnetic field decay
rate.  This observation motivates using a coordinate system that is
regular on the event horizon and extends smoothly into the black hole
interior.

\subsection{Maximal Slicing: Interior and Exterior Solution}
\label{sec7.3}

\begin{figure*}[t]
\begin{center}
\leavevmode
\epsfxsize=5.5in
\epsffile{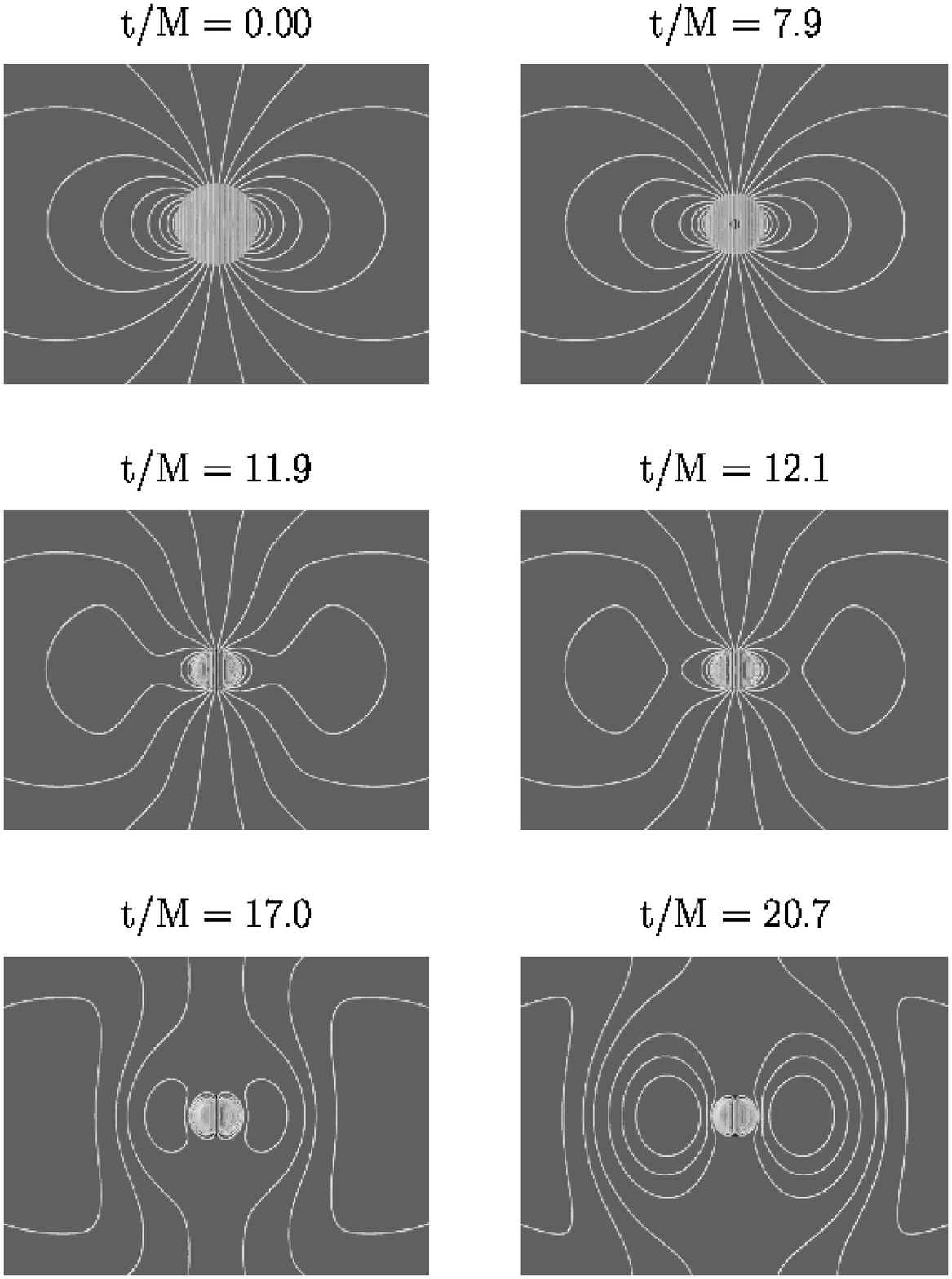}
\end{center}
\caption{Snapshots of the interior and exterior magnetic field lines 
at select {\it maximal} time slices for the same collapse depicted in
Figure \ref{fig1}. The star again collapses from rest at an initial areal
radius $R_s(0) = 4 M$. Points are plotted in a meridional plane
using areal radii. The light shaded sphere covers the matter interior;
the black shaded sphere covers the region inside the event horizon. 
Soon after the last snapshot at $t = 20.7$ grid stretching causes the
quality of the numerical integration to deteriorate in this gauge. 
See text for details.} 
\label{fig4}
\end{figure*}

\begin{figure*}[t]
\begin{center}
\leavevmode
\epsfxsize=5.5in
\epsffile{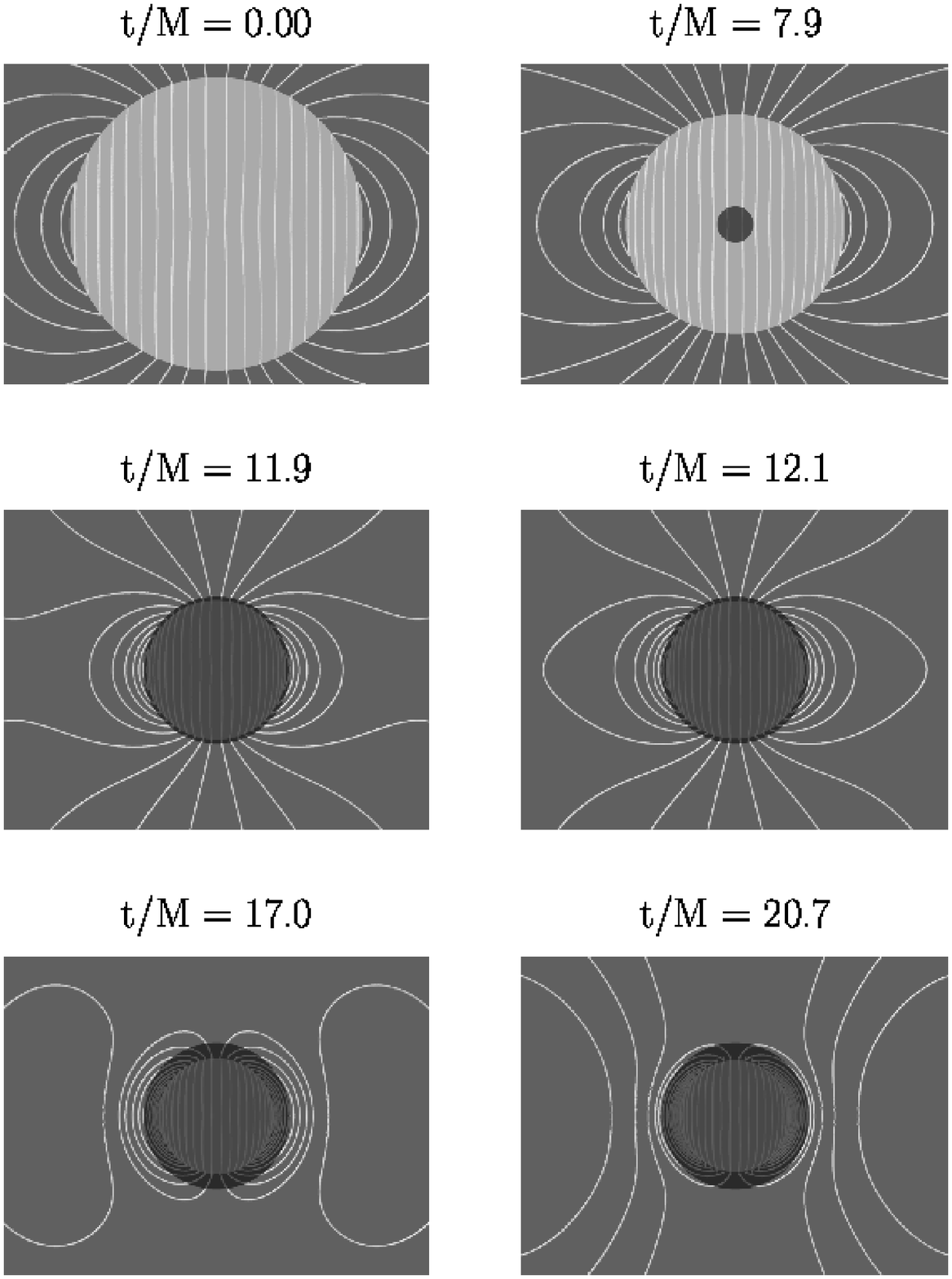}
\end{center}
\caption{ Blow-up of the same collapse depicted in Figure \ref{fig4}.
Note how in this gauge the stellar surface approaches a limit surface at
$r_s \approx 1.5 M$ at late times. The horizon grows to its final
value $r_s = 2M$ once all the matter crosses inside this radius.}
\label{fig5}
\end{figure*}

We now follow the same collapse but adopt maximal time slicing 
and isotropic (minimal shear) spatial coordinates.
This coordinate system extends regularly into the 
black hole.  Maximal slicing is imposed by requiring the trace of 
the extrinsic curvature to vanish at all times:
\begin{equation}
K=0=\frac{dK}{dt}.
\end{equation}
Taking the trace of equation (I.2.8) then yields an elliptic equation
for $\alpha$, which in general has to be solved numerically.  Maximal
slices of Oppenheimer-Snyder spacetime in isotropic coordinates may be
determined by solving {\it ordinary} differential equations. Solutions
are presented in Petrich, Shapiro \& Teukolsky (1985) and provide
$\alpha$, $\beta$ and $A$, which we substitute in equations
(\ref{edot1}) and (\ref{adot1}) for the evolution of $e$ and $a$.
These coordinates cover both the interior and the exterior of the
star, so that we can map out the interior as well as the exterior
field on these slices.

The Lorentz factor between a normal observer and a comoving observer at
the stellar surface $\gamma = - u_{a} n^{a}$ can be computed from the
numerical solution presented in Petrich, Shapiro \& Teukolsky (1985)
(i.e., $\gamma = 1/(1-T^2)$ and $v=|T|$, where $T$ is given by
equation (6.11) of that paper).  This is the
velocity that is needed in the boundary condition
(\ref{e_inner_condition}).

We numerically solve equations (\ref{edot1}) and (\ref{adot1}) on
maximal slices using implicit finite difference equations that are
identical in form to those used on Schwarzschild slices (see Appendix
\ref{secB}).

In Figures \ref{fig4} and \ref{fig5} we show snapshots of the collapse
in maximal slicing.  Unlike Schwarzschild slices and the Kerr-Schild
slices below, maximal slicing also covers the matter interior, which
allows us to trace the interior magnetic field lines as well as locate
the apparent and event horizons in these plots.  We can also identify
the limit surface at $R_s \approx 1.5 M$ which the matter surface
approaches at late times (Petrich, Shapiro \& Teukolsky 1985). Note
that in the interior, a field line must connect the same Lagrangian
fluid element at all times to satisfy the ``frozen-in'' requirement of
MHD. This criterion provides a check on our field line tracer. For
comoving observers in the interior, the electric field vanishes, but
as viewed by a normal observer in maximal slicing the electric field
is non-vanishing in general (see Section \ref{sec5.2.1} for the
transformation).

In maximal slicing the metric is not static, so that we cannot simply
apply the electromagnetic energy conservation check developed in
Appendix \ref{secA}.

Maximal slicing has some very desirable properties, but at late times
grid stretching effects make the computation even of the background
models increasingly difficult (Shapiro \& Teukolsky 1986).  We
terminated the calculation shortly after $t = 20M$.

\subsection{Kerr-Schild Slicing: Black Hole Excision}
\label{sec7.4}

The problems of maximal slicing can be avoided by using Kerr-Schild
(or ingoing Eddington-Finkelstein) coordinates, which are analytic but
smoothly extend across the horizon.  This slicing is not ``singularity
avoiding'', so that it requires that the interior of a black hole be
excised from the numerical grid.  Adopting horizon-penetrating
coordinate systems together with black hole excision is currently the
most promising approach to dynamical simulations of black holes.  In
this calculation we apply this approach to the collapse of a
magnetized star.  Since the metric in Kerr-Schild coordinates
(equation (\ref{metric_ks}) below) takes a form that is different from
(\ref{metric_isotropic}), several parts of the above calculations,
including the form of Maxwell's equations, boundary conditions and
initial data, have to be modified.

\subsubsection{Kerr-Schild coordinates}
\label{sec7.4.1}

In Kerr-Schild coordinates, the metric takes the form
\begin{equation} \label{metric_ks}
ds^2 = - (1 - \frac{2M}{r_s}) dt^2 + \frac{4M}{r_s} dt dr_s	
	+ (1 + \frac{2M}{r_s}) dr_s^2 + r_s^2 d\Omega^2,
\end{equation}
where $r_s$ is the Schwarzschild (or areal) radius.  From the metric we
can identify the lapse as
\begin{equation}
\alpha = \left( \frac{r_s}{r_s+2M} \right)^{1/2},
\end{equation}
the radial shift as 
\begin{equation} \label{shift_ks}
\beta = \frac{2M}{r_s + 2M},
\end{equation}
and the diagonal components of the spatial metric $\gamma_{ij}$ as
\begin{equation}
\gamma_{r_s r_s} = 1 + \frac{2M}{r_s}
\end{equation} 
and 
\begin{equation}
\gamma_{\theta\theta} = \frac{1}{\sin^2 \theta} \gamma_{\phi\phi} = r_s^2
\end{equation}
(see, e.g., Lehner et al. 2000).  The extrinsic curvature can then be
computed from equation (I.2.9), which yields
\begin{equation}
K_{r_s r_s} = - \frac{2M}{r_s^3} (r_s + M) \alpha
\end{equation}
and 
\begin{equation}
K_{\theta\theta} = \frac{1}{\sin^2 \theta} K_{\phi\phi} = 2 M \alpha.
\end{equation}
The trace of the extrinsic curvature is
\begin{equation} \label{K_ks}
K = K^i_{~i} = \frac{2 M}{r_s^2} \, \frac{1 + 3M/r_s}{(1 + 2M/r_s)^{3/2}}
	= \frac{2M\alpha}{r_s^2} \, \frac{r_s + 3M}{r_s + 2M}.
\end{equation}

\subsubsection{Radial geodesics in Kerr-Schild coordinates}

\begin{figure*}[t]
\begin{center}
\leavevmode
\epsfxsize=3in
\epsffile{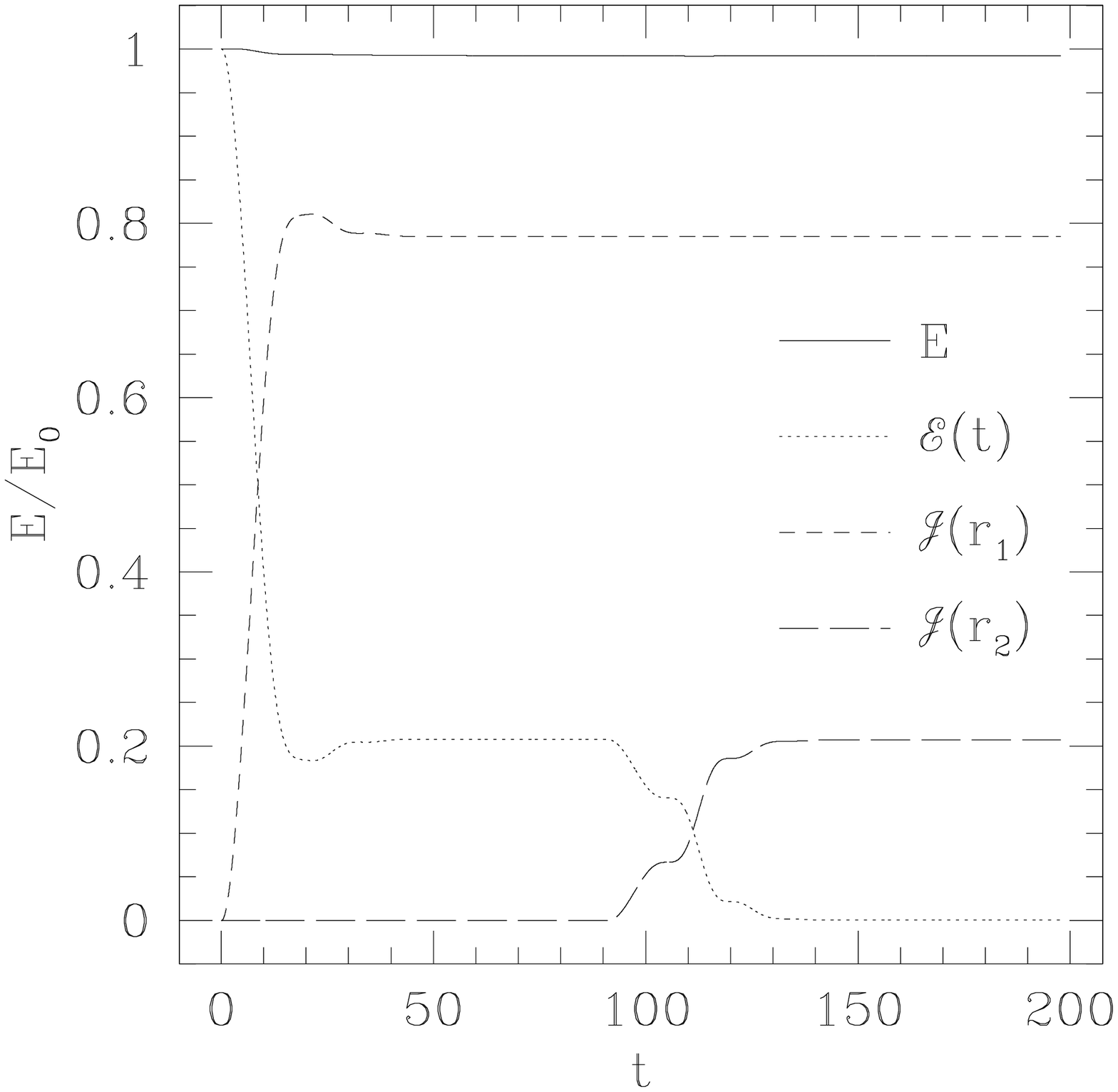}
\end{center}
\caption{Electromagnetic energy conservation in Kerr-Schild time
slicing.  The outer boundary is located at $R_{\rm out} = 100 M$, the
exterior of the star is covered with 5000 gridpoints, and the Courant
factor is 0.5. The solid line denotes the total energy $E$, which
should be conserved (see equation ~(\ref{Econs})).  The dotted line is
the energy ${\mathcal E}(t)$ contained between the radii $r_1 = 1.01
R_s(0) = 4.04 M$ and $r_2 = 80 M$; the short dashed line is the
integrated flux across $r_1$, ${\mathcal J}(r_1)$, and the long dashed
line is the integrated flux across $r_2$, ${\mathcal J}(r_2)$.  Note
that approximately 80\% of the energy enters the black hole, while
only 20\% is radiated away to infinity.}
\label{fig6}
\end{figure*}

The surface of the collapsing star follows a radial geodesic, which
can be determined as follows.  The areal radius as a function of
proper time has to satisfy the same equations,
\begin{equation}
\begin{array}{rcl}
R_s & = & \displaystyle \frac{R_s(0)}{2} (1 + \cos \eta) \\
\tau & = & \displaystyle 
	\left( \frac{R_s^3(0)}{8M} \right)^{1/2} 
	( \eta + \sin \eta)
\end{array}
\end{equation}
as before (see Section \ref{sec3}), since both quantities are gauge
invariant parameters.  Here $R_s(0)$ is the initial Schwarzschild
radius at which the surface is momentarily at rest.  An equation for
the Kerr-Schild coordinate time $t$ can be derived from the
conservation of $p_t$
\begin{equation}
p_t = g_{tt} p^t + g_{t r_s} p^{r_s} = \mbox{const} \equiv - E,
\end{equation}
or
\begin{equation} \label{dtdtau1}
\left( 1 - \frac{2M}{r_s} \right) \frac{dt}{d\tau} 
	- \frac{2M}{r_s} \frac{dr_s}{d\tau} = \tilde E.
\end{equation}
Combining this equation with the normalization condition for the
four-velocity, $u_{a} u^{a} = -1$, yields
\begin{equation}
\frac{dr_s}{d\tau} = - \left(\tilde E^2 - 1 + \frac{2M}{r_s} \right)^{1/2}.
\end{equation}
Evaluating this result for the surface at $\tau = 0$, when $dR_s/d\tau
= 0$, determines the energy 
\begin{equation}
\tilde E = \left( 1 - \frac{2M}{R_s(0)} \right)^{1/2}.
\end{equation}
Inserting the last two equations into (\ref{dtdtau1}) then yields
\begin{equation} \label{dtdtau2}
\frac{dt}{d\tau} = \frac{(1 - 2M/R_s(0))^{1/2}}{1 - 2M/R_s} 
	- \frac{2M}{R_s} \, \frac{(2M/R_s - 2M/R_s(0))^{1/2}}{1-2M/R_s}.
\end{equation}
Interestingly, this equation can be integrated analytically to yield
\begin{eqnarray}
& & t = 2M \left(\frac{R_s(0)}{2M} - 1 \right)^{1/2} 
	\left(\eta + \frac{R_s(0)}{4M} (\eta + \sin \eta) \right) \\
& & + 2 M \ln \left( 
	\frac{1 + \frac{R_s}{2M} - \frac{2R_s}{R_s(0)} + 
	2 \sqrt{(\frac{R_s}{2M} - \frac{R_s}{R_s(0)})
	(1 - \frac{R_s}{R_s(0)})}}
		{\frac{R_s(0)}{2M} - 1} \right) \nonumber
\end{eqnarray}	
The coordinate time $t_f$ for infall from $r_s = R_s(0) > 2M $ to the
singularity at $r_s = 0$ is therefore
\begin{equation}
t_{f} = 2 \pi M \left(1 + \frac{R_s(0)}{4M} \right) 
	\left(\frac{R_s(0)}{2M} - 1 \right)^{1/2} 
	- 2 M \ln \left(\frac{R_s(0)}{2M} - 1 \right).
\end{equation}
This time is evidently finite, which reflects the fact that
Kerr-Schild coordinates are horizon penetrating.  This also
demonstrates the necessity of excising the black hole interior, since
otherwise the central singularity would be encountered after $t =
t_f$.

The Lorentz factor corresponding to a boost from an observer comoving
with the stellar surface to a local normal observer can be found from
\begin{equation} \label{gamma_ks}
\gamma = - n_{a} u^{a} = \alpha u^t = 
         \left(1 - \frac{4M^2}{R_s^2} \right)^{-1/2}.
\end{equation}
The relative velocity  $v$ of the normal observer as measured by a comoving
observer at the surface is then given by
\begin{equation} \label{v_ks}
v = - \frac{2M}{R_s}.
\end{equation}
A minus sign appears in equation~(\ref{v_ks}) because the shift drives
the radial velocity of normal observer inward more rapidly than the
surface.
\begin{figure*}[t]
\begin{center}
\leavevmode
\epsfxsize=5.5in
\epsffile{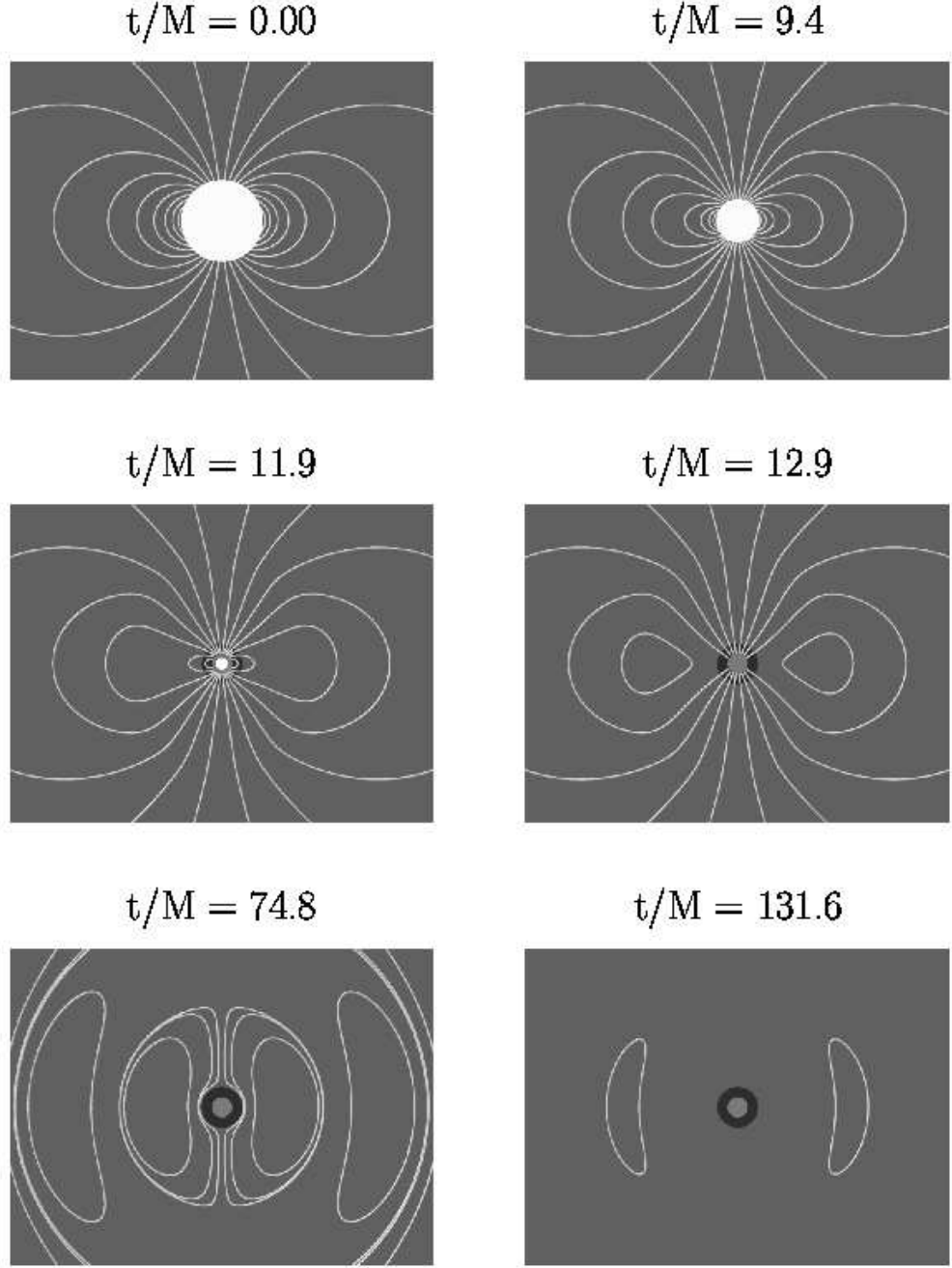}
\end{center}
\caption{Snapshots of the exterior magnetic field lines at select {\it
Kerr-Schild} time slices for the same collapse depicted in Figure
\ref{fig1}. The star again collapses from rest at an initial areal
radius $R_s(0) = 4 M$. Points are plotted in a meridional
plane using areal radii.  The white shaded sphere covers the matter
interior; the black shaded area covers the region inside the event
horizon; the grey shaded area covers the region inside $r_s = M$
excised from the numerical grid once the matter passes inside. In this
gauge, using excision, we are able to integrate reliably to arbitrary
late times.  See text for details.}
\label{fig7}
\end{figure*}

\begin{figure*}[t]
\begin{center}
\leavevmode
\epsfxsize=5.5in
\epsffile{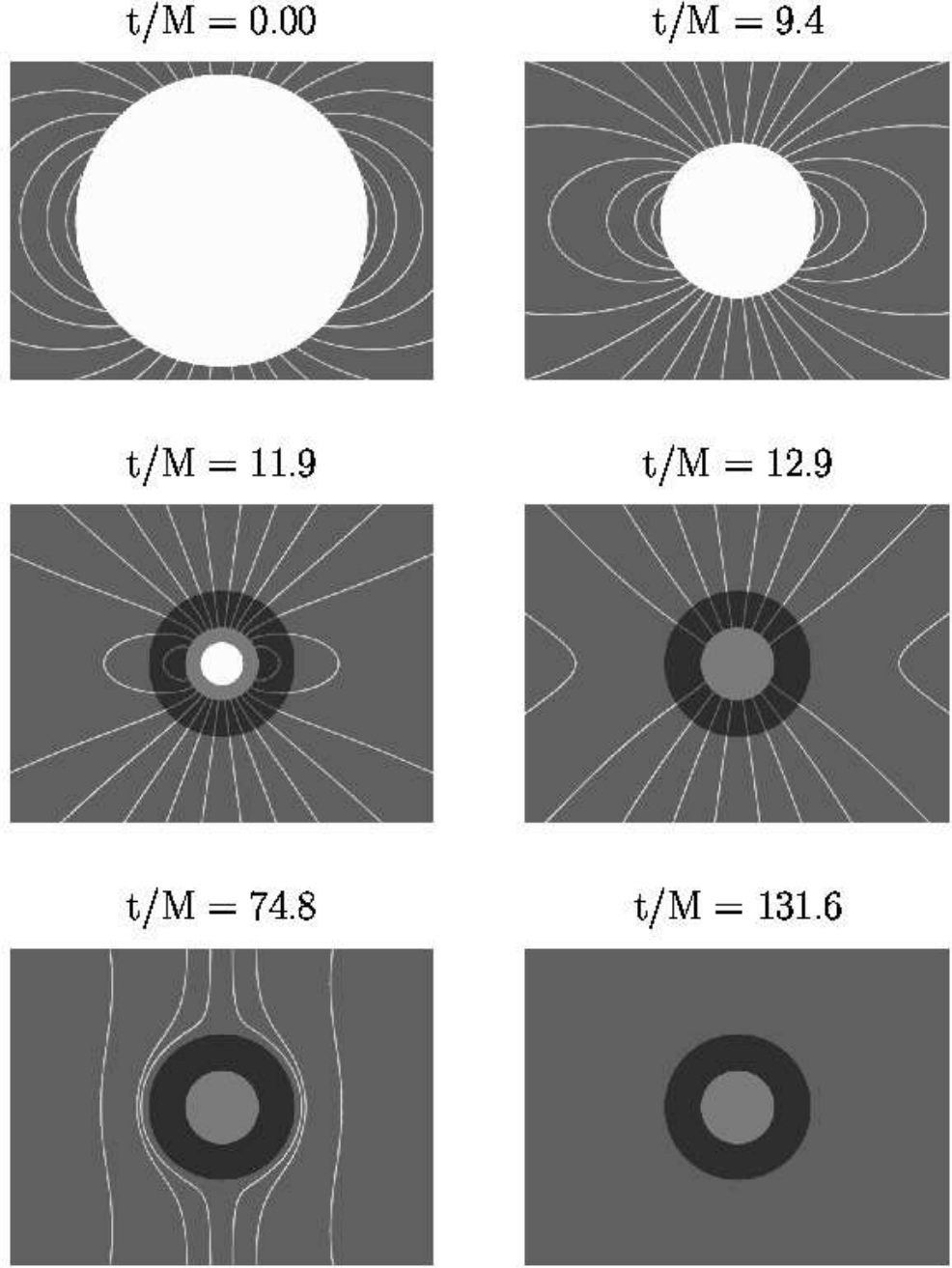}
\end{center}
\caption{ Blow-up of the same collapse depicted in Figure \ref{fig7}.
By the end of the integration, all exterior electromagnetic fields in
the vicinity of the black hole have been captured or radiated away.}
\label{fig8}
\end{figure*}

\begin{figure*}[t]
\begin{center}
\leavevmode
\epsfxsize=5.5in
\epsffile{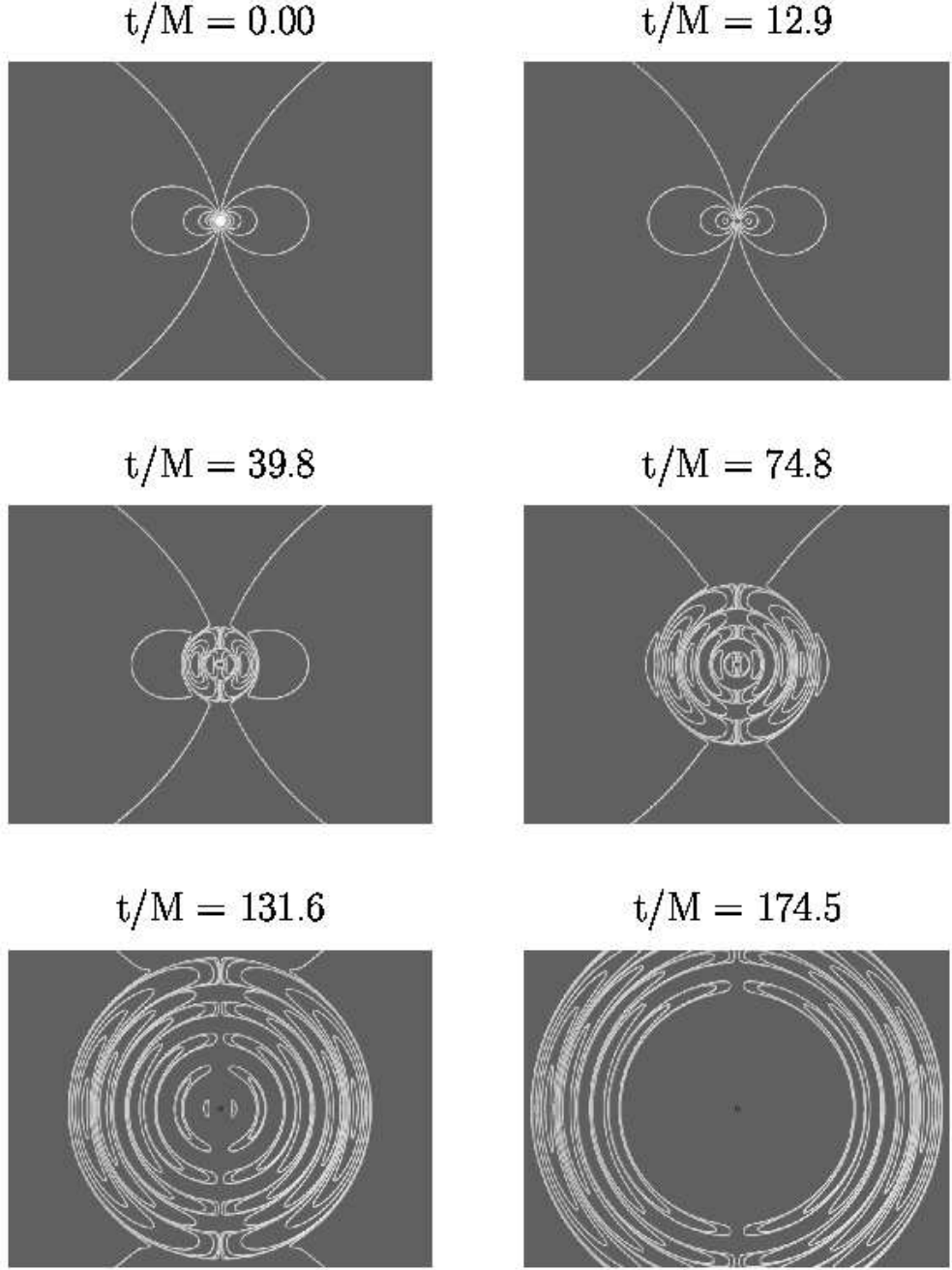}
\end{center}
\caption{ Far-zone view of the same same collapse depicted in Figures
\ref{fig7} and \ref{fig8}, but now extending to very late times. Note
the transformation of the dipole magnetic field from a quasi-static
longitudinal to a transverse electromagnetic wave.}
\label{fig9}
\end{figure*}

\subsubsection{Exterior initial data in Kerr-Schild coordinates}
\label{sec7.4.3}

Given the boost parameters $\gamma$ and $v$, we can transform the
magnetic dipole initial data of Section \ref{sec6.2} into Kerr-Schild
coordinates.  Applying the transformation rules (\ref{EBtrans}) we
find
\begin{equation}
\begin{array}{rcl}
B^{\hat r}_{\rm KS} & = & B^{\hat r}_{\rm WS} \\
B^{\hat \theta}_{\rm KS} & = & \gamma B^{\hat \theta}_{\rm WS} \\
B^{\hat \phi}_{\rm KS} & = & 0 \\
E^{\hat r}_{\rm KS} & = & E^{\hat r}_{\rm WS} = 0 \\
E^{\hat \theta}_{\rm KS} & = & 
	\gamma (E^{\hat \theta}_{\rm WS} - v B^{\hat \phi}_{\rm WS}) = 0\\
E^{\hat \phi}_{\rm KS} & = & 
	\gamma (E^{\hat \phi}_{\rm WS} + v B^{\hat \theta}_{\rm WS}) = 
	\gamma v B^{\hat \theta}_{\rm WS}.
\end{array}
\end{equation}
Here the subscript KS refers to Kerr-Schild data, and WS to the
Wasserman-Shapiro exterior magnetic dipole solution presented in
Section \ref{sec6.2}.  We note that even at $t=0$ the electric field
is non-vanishing as seen by a normal observer in Kerr-Schild
coordinates, since a normal observer is moving with respect to a
static observer.

\subsubsection{Maxwell's equations in Kerr-Schild coordinates}

\begin{figure*}[t]
\begin{center}
\leavevmode
\epsfxsize=3in
\epsffile{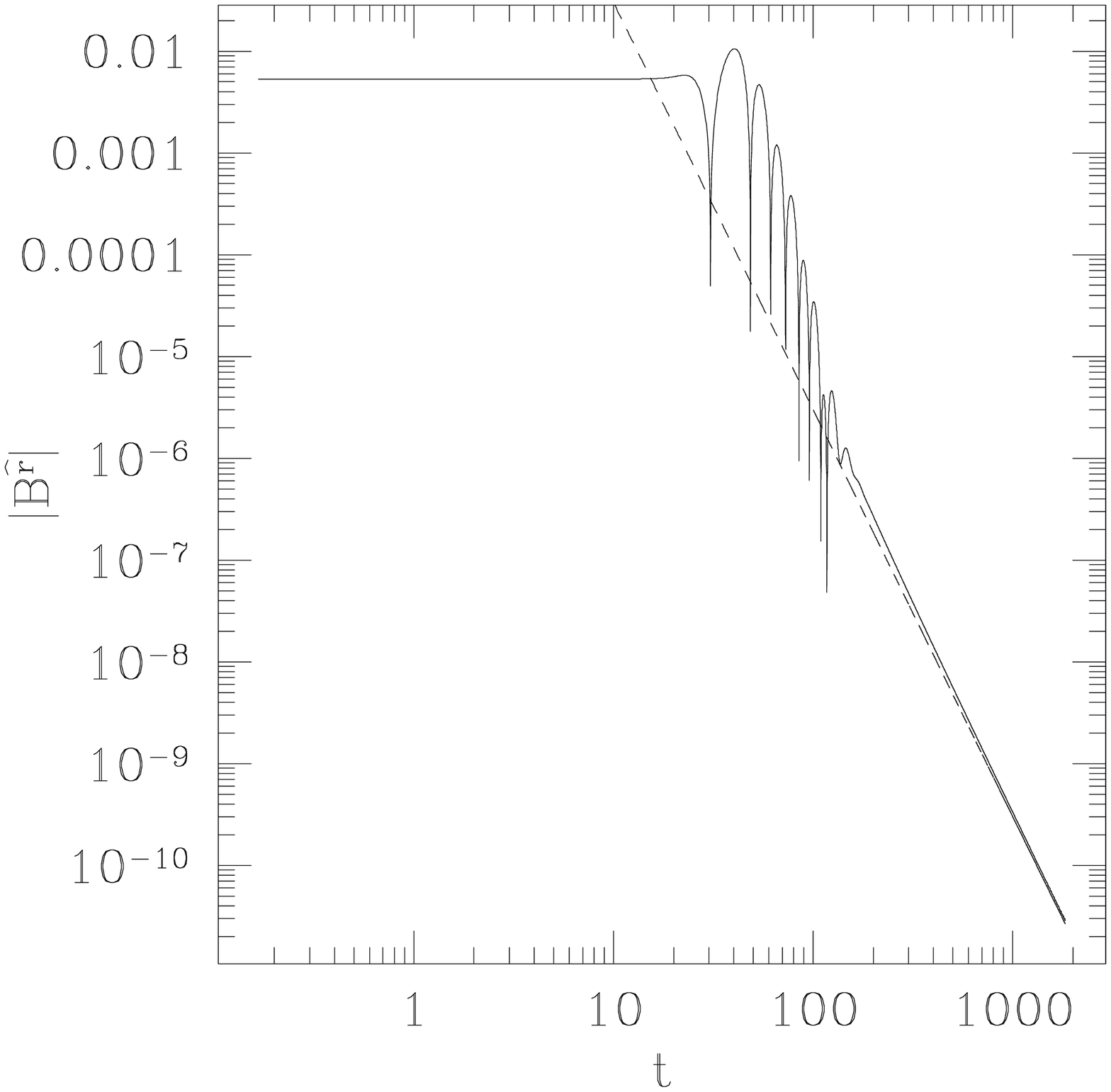}
\end{center}
\caption{The absolute value of the radial magnetic field $|B^{\hat
r}|$ at $r = 20M$ as a function of time.  For this simulation, the
outer boundary was placed at $2000M$, the numerical grid consisted of
6000 radial gridpoints, and the Courant factor was 0.5.  The dashed
line is a power law $t^{-4}$, which approximates the late-time
fall-off extremely well.}
\label{fig10}
\end{figure*}

\begin{figure*}[t]
\begin{center}
\leavevmode
\epsfxsize=3in
\epsffile{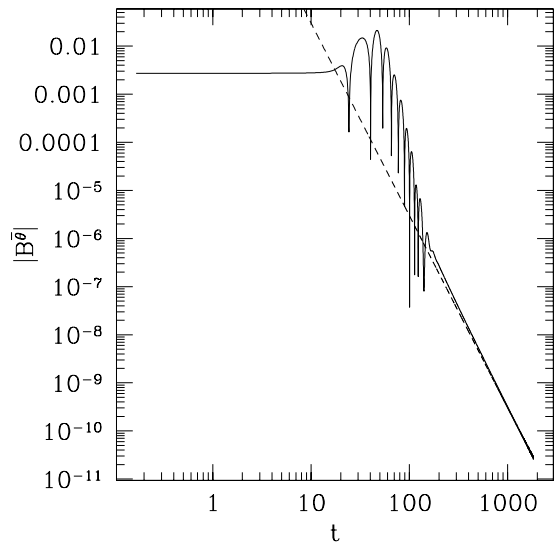}
\end{center}
\caption{Same as Fig.~\ref{fig10}, except for  $|B^{\hat \theta}|$.}
\label{fig11}
\end{figure*}

\begin{figure*}[t]
\begin{center}
\leavevmode
\epsfxsize=3in
\epsffile{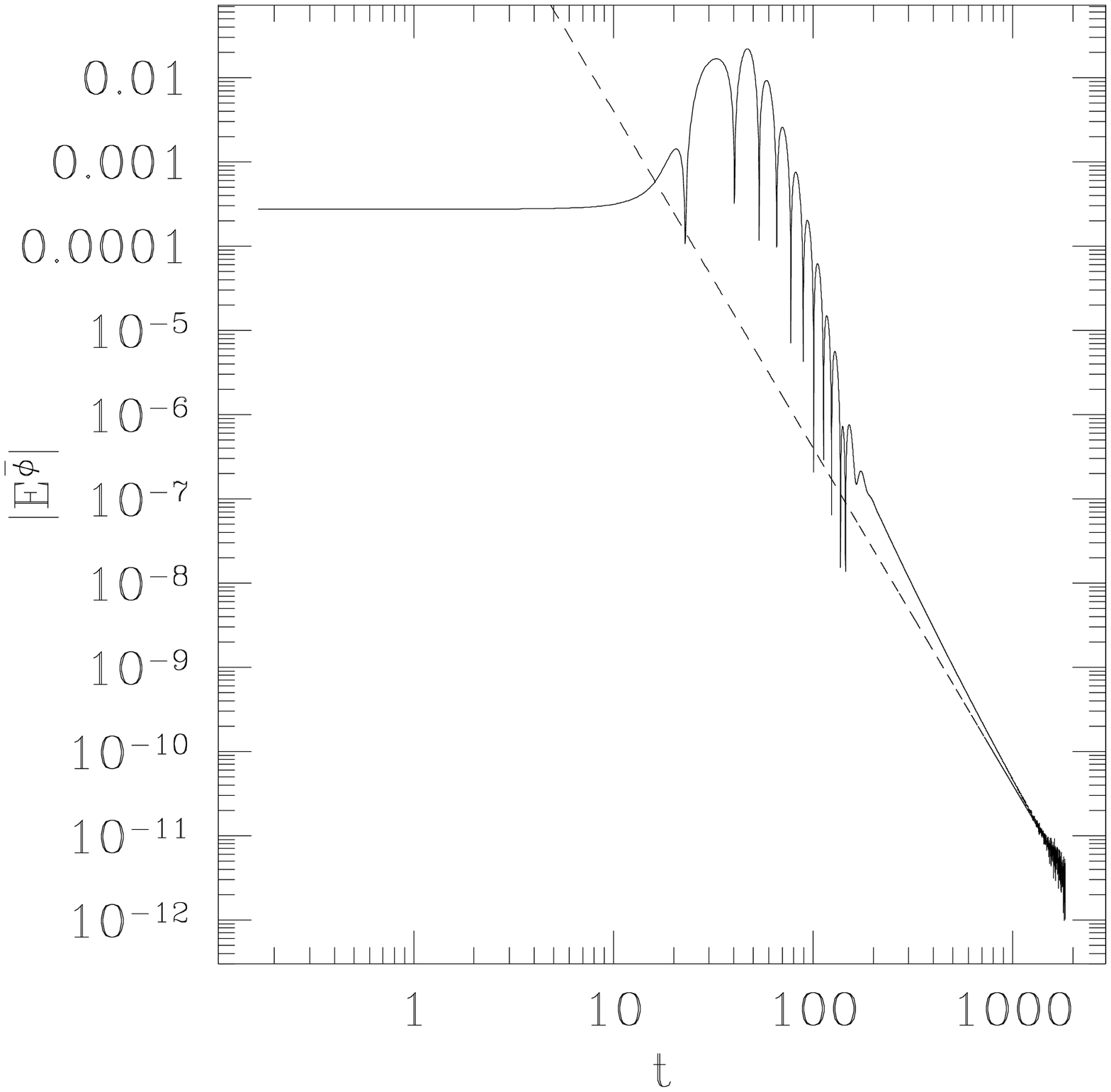} 
\end{center}
\caption{
Same as Fig.~\ref{fig10}, except for  $|E^{\hat \phi}|$.}
\label{fig12}
\end{figure*}

In Section \ref{sec5.1} we evaluated Maxwell's equations (\ref{Adot1})
and (\ref{Edot1}) for a metric of the form (\ref{metric_isotropic}).
Since the Kerr-Schild metric (\ref{metric_ks}) has a slightly
different form, we have to re-evaluate Maxwell's equations.

Our derivation closely follows that of Section \ref{sec5.1}.  We
insert the metric coefficients of Section \ref{sec7.4.1} into
equations (\ref{Adot1}) and (\ref{Edot1}), choose the Coulomb gauge
condition by setting $\Phi = 0$, ensure axisymmetric fields by
assuming that $A_{\phi} = A_{\phi}(t,r,\theta)$ and $E^{\phi} =
E^{\phi}(t,r,\theta)$ are the only non-vanishing components, and
finally use the expansions (\ref{A_expansion}) and
(\ref{E_expansion}).  The resulting equations for $e$ and $a$ are
\begin{equation} \label{edot_ks}
\partial_t e = \alpha \left(\alpha^2 a_{,r_s}\right)_{,r_s} 
	- \alpha \, \frac{l(l+1)}{r_s^2} \, a + \alpha K e 
	+ \beta r_s^2 \left( \frac{e}{r_s^2} \right)_{,r_s}
\end{equation}
and
\begin{equation} \label{adot_ks}
\partial_t a = \alpha e + \beta a_{,r_s},
\end{equation}
where the metric coefficients $\alpha$ and $\beta$ and the trace of
the extrinsic curvature $K$ are given in Section \ref{sec7.4.1}.  

In terms of $e$ and $a$, the components of the electric and magnetic
fields are given by
\begin{eqnarray}
E^{\hat \phi} & = & \frac{\sin \theta}{r_s} \, e \\
B^{\hat r} & = & - \frac{2 \cos \theta}{r_s^2} \, a \\
B^{\hat \theta} & = & \frac{\sin \theta}{(1 + 2M/r_s)^{1/2} r_s} \, a_{,r_s}.
\end{eqnarray}

\subsubsection{Boundary Conditions}

For the early part of the evolution, the inner boundary conditions
result from junction conditions on the electromagnetic fields at the
stellar surface.  The derivation of these conditions follows that of
Section \ref{sec5.2.1} and results in
\begin{equation} \label{a_inner_ks}
a = - \frac{r_s^2 B_{\hat r}^{(in)}}{2 \cos \theta}
\end{equation}
and
\begin{equation} \label{e_inner_ks}
e = \frac{v}{(1 + 2M/r_s)^{1/2}} \, a_{,r_s}.
\end{equation}

Once the stellar surface has passed through some radius $R_{\rm ex} <
2M$ inside the horizon, we switch from the junction condition on the
stellar surface to an excision boundary condition at $R_{\rm ex}$.  We
typically set $R_{\rm ex} = M$, but the results are insensitive to the
choice provided $R_{\rm ex}$ is well inside the horizon at $2M$.  Once
we have switched to an excision boundary condition, we fix the
numerical grid so that $R_{\rm ex}$ is located half-way between the
first two gridpoints.

We have experimented with a number of different excision boundary
conditions, but found, as expected, that the exact implementation had
hardly any effect on the results in the exterior of the black hole, as
long as $R_{\rm ex}$ is far enough inside the event horizon.
Following a similar approach by Alcubierre \& Br\"ugmann (2001) for
dynamical black hole simulations, we settled on simply copying the
values of $e$ and $a$ from the second gridpoint to the first at every
timestep, which is equivalent to imposing
\begin{equation} \label{inner_ex}
e_{,r_s} = a_{,r_s} = 0
\end{equation}
(see equations (\ref{B_inner_bc_a}) and (\ref{B_inner_bc_b})).

\subsubsection{Numerical results}

We have implemented the evolution equations (\ref{edot_ks}) and
(\ref{adot_ks}) using both implicit and explicit finite difference
methods (see Appendix \ref{secB} for the finite difference equations).
The results presented here were obtained with the explicit scheme.

Since the Kerr-Schild metric (\ref{metric_ks}) is independent of time,
we can calibrate the performance of the code with the same global
electromagnetic energy conservation test as we did in Schwarzschild
coordinates (Section \ref{sec7.2}).  The necessary expressions can be
found in Appendix \ref{secA}.  We show results for the energy
conservation check in Figure \ref{fig6}.  It is quite obvious that
this integration works much better than that in Schwarzschild
coordinates (cf. Figure \ref{fig3}), due to regularity across the
horizon. We can follow the electromagnetic fields to arbitrary late
times, and can track the late-time behavior reliably.

Figure \ref{fig6} also shows that about 80\% of the energy contained
in the initial magnetic field is absorbed by the black hole, while
about 20\% of the radiation is emitted in an electromagnetic
wavepacket.

In Figures \ref{fig7} through \ref{fig9} we show snapshots of magnetic
field lines for a star collapsing from $R_s(0) = 4M$ as viewed in
three different regions.  It can be seen very clearly how the
collapse of the magnetized star emits a pulse of electromagnetic
radiation, which travels to infinity at the speed of light.  Similar
behavior is seen in plots of the electric field at late times.

We can now evaluate the late-time fall-off of the electromagnetic
fields.  In Figures \ref{fig10} through \ref{fig12} we show the values
of $|B^{\hat r}|$, $|B^{\hat \theta}|$ and $|E^{\hat \phi}|$ at $r =
20M$ as a function of time.  As expected, the fields are nearly
constant up to $t = 20M$. Once the collapse gets underway, the
exterior longitudinal field starts to evolve to match the increasing
interior frozen-in field. Soon after $t=20M$, the first
electromagnetic pulse induced by the collapse passes the observer at
$r = 20M$.  Sign changes in the field appear as down-ward spikes in
the plots.  After $t \sim 100M$, the exterior fields are predominantly
characterized by a decaying dipole with a
power-law fall-off with time.  In Figures \ref{fig10} through
\ref{fig12} we have plotted the power-laws curves varying like
$t^{-4}$ to which the fields approach for late times.  This finding
for the exterior fields agrees with the general results of Price
(1972a,b; cf.~Thorne, 1971) who showed that multipole electromagnetic
fields scattering off a static Schwarzschild black hole should decay
as $t^{-(2l+2)}$.

Measurement of the wavelength of the late-time outgoing
electromagnetic waves provides further verification of our numerical
integrations. We find that the wavelength satisfies $\lambda \approx
24.7M$. This is the value of a quasi-normal mode for electromagnetic
dipole radiation scattering off a static Schwarzschild black hole
(Ferrari \& Mashhoon 1984).  This finding shows that the late-time
radiation is ringing radiation that has a frequency determined by the
mass of the hole and is independent of the initial electromagnetic
field profile.

\section{Summary}
\label{sec8}

We have solved for the evolution in general relativity of the
electromagnetic field of a magnetic star that collapses from rest to a
Schwarzschild black hole. We adopted a ``dynamical Cowling'' model by
which the matter and gravitational field are prescribed by the
Oppenheimer-Snyder solution for spherical collapse. Starting from
Maxwell's equations in $3+1$ form, we solved for the magnetic and
electric fields both in the vacuum exterior and the stellar interior.
We assumed that the initial star was threaded by a dipole magnetic
field and that the interior behaves as an MHD fluid. We showed how the
electromagnetic junction conditions could be applied at the stellar
surface in order that the fields in the MHD interior, which were
determined analytically, could be matched to the fields in the vacuum
exterior. The exterior fields were evolved numerically.

To gain experience with solving Maxwell's equations numerically in a
dynamical spacetime containing both matter and a black hole, we solved
the above problem in three different coordinate gauges: Schwarzschild,
maximal and Kerr-Schild. While the first two coordinate systems have
singularity avoiding properties, they both lead to growing numerical
inaccuracies in the exterior as the surface approaches the horizon (in
the case of Schwarzschild time slicing) or a limit surface inside the
horizon (in the case of maximal slicing) at late times.  Kerr-Schild
slicing is horizon penetrating and does not suffer from grid
stretching; accordingly this slicing enables us to apply excision
boundary conditions once the star collapses inside the horizon.
Adopting Kerr-Schild slicing coordinates with black hole excision
allowed us to integrate Maxwell's equations to arbitrary late
times. We followed the transition in the exterior of a quasi-static,
longitudinal magnetic field, which evolved during the collapse in
to match the growing frozen-in interior field, to a transverse
electromagnetic wave. By the end of the simulation, the
electromagnetic field energy in the near zone outside the black hole
had either been captured by the hole or radiated away to large
distances, in accord with the ``no-hair theorem''. At late times we
recovered the expected power-law decay of the dipole fields with time,
as well as the expected wavelength of the ringdown radiation.

There exist a number of very important astrophysical problems which
require the solution of the coupled Maxwell-Einstein-MHD equations in
a strong, dynamical gravitational field. For example, learning the
outcome of rotating stellar core or supermassive star collapse, the
mechanism for gamma-ray bursts, and/or the fate of the remnant of a
binary neutron star merger may all require solving this coupled system
of equations. The solution we have presented for our simple collapse
scenario should serve as a preliminary guide to the design and testing
of more sophisticated codes capable of handling these more complicated
scenarios.

\acknowledgments

It is a pleasure to thank M. Duez, C. Gammie, and H.--J. Yo for useful
discussions and the Undergraduate Research Team (H. Agarwal,
R. Cooper, B. Hagan and D. Webber) at the University of Illinois at
Urbana-Champaign (UIUC) for assistance with visualization.  This work
was supported in part by NSF Grants PHY-0090310 and PHY-0205155 and
NASA Grant NAG5-10781 at UIUC and NSF Grant PHY 0139907 at Bowdoin
College.

\begin{appendix}

\section{Global Electromagnetic Energy Conservation}
\label{secA}

\subsection{General Discussion}
\label{secA.1}

The conservation of energy can be verified very easily if the exterior
metric is independent of time, so that 
\begin{equation}
{\bf \xi} = \frac{\partial}{\partial t}
\end{equation}
is a Killing vector.  In that case, 
\begin{equation}
J_{\mu} \equiv T_{\mu\nu} \xi^{\nu}
\end{equation}
is a conserved current and satisfies
\begin{equation}
\oint_{\partial V} J^{\alpha} d^3 \Sigma_{\alpha} = 0,
\end{equation}
where ${\partial V}$ denotes the 3-surface enclosing the 4-volume $V$.
If the volume $V$ is confined by the radii $r_1$ and $r_2$ and
the times $t_1$ and $t_2$, the conservation law can be written as
\begin{eqnarray}
 & \displaystyle 
\left.\int_{r=r_1}^{r_2} \int_{\theta=0}^{\pi} \int_{\phi=0}^{2\pi} 
J^0 \sqrt{-g} dr d\theta d\phi \right|_{t_1}^{t_2} & \nonumber \\
& \displaystyle
 ~~~= - \left. \int_{t=t_1}^{t_2} \int_{\theta=0}^{\pi} \int_{\phi=0}^{2\pi} 
J^r \sqrt{-g} dt d\theta d\phi \right|_{r_1}^{r_2} & .
\end{eqnarray}
Defining the energy contained between $r_1$ and $r_2$ at time $t$ as
\begin{equation} \label{AE}
{\mathcal E}(t) \equiv 
\int_{r=r_1}^{r_2} \int_{\theta=0}^{\pi} \int_{\phi=0}^{2\pi} 
J^0 \sqrt{-g} dr d\theta d\phi
\end{equation}
and the integrated flux across the radius $r$ as
\begin{equation} \label{AJ}
{\mathcal J}(r) \equiv 
\int_{t=t_1}^{t_2} \int_{\theta=0}^{\pi} \int_{\phi=0}^{2\pi} 
J^r \sqrt{-g} dt d\theta d\phi
\end{equation}
the conservation law can be rewritten as
\begin{equation} \label{Acons}
{\mathcal E}(t_2) - {\mathcal E}(t_1) = {\mathcal J}(r_2) - {\mathcal J}(r_1).
\end{equation}
The meaning of this statement is quite evident: the difference in the
energy contained between $r_1$ and $r_2$ at two different times has to
equal the net energy flux entering the region, integrated over the
time interval.  An alternative way of writing the conservation law is
\begin{equation} \label{Econs}
E/E_0 = \left ({\mathcal E}(t_2) - {\mathcal J}(r_2) 
     + {\mathcal J}(r_1) \right)/E_0 = 1,
\end{equation}
where $E_0 = {\mathcal E}(t_1)$ is the initial energy between $r_1$
and $r_2$.  We use equation~(\ref{Econs}) in plotting our results in
the text.

Since the exterior metric is independent of time in both Schwarzschild
coordinates (Section \ref{sec7.2}) and Kerr-Schild coordinates
(Section \ref{sec7.4}), energy conservation can be verified quite
easily in both of those coordinate systems.  To do so, the energy
(\ref{AE}) and the flux (\ref{AJ}) have to be expressed in terms of
the relevant metric quantities, which we do separately in the two
following Sections.  A similar analysis was used in Appendix A of
Pavlidou, Tassis, Baumgarte \& Shapiro (2000) for the energy
conservation of a scalar field in the static spacetime of a spherical,
equilibrium star.

\subsection{Schwarzschild Slicing}
\label{secA.2}

In isotropic Schwarzschild coordinates the metric takes the form
(\ref{metric_isotropic}) with $\alpha$, $A$ and $\beta$ given by
(\ref{alpha_SS}), (\ref{A_SS}) and (\ref{beta_SS}).  The determinant
of the metric is
\begin{equation}
\sqrt{-g} = \alpha A^3 r^2 \sin \theta,
\end{equation}
and the fluxes $J^0$ and $J^r$ in (\ref{AE}) and (\ref{AJ}) can be 
expressed as
\begin{equation}
J^0 = T^0_{~0} =  T^{\hat 0}_{~\hat 0} = - T^{\hat 0 \hat 0}
	= - \frac{1}{8 \pi}({\bf E}^2 + {\bf B}^2)
\end{equation}
and
\begin{equation}
J^r =  T^r_{~0} = \frac{\alpha}{A}  T^{\hat r}_{~\hat 0} 
	= - \frac{\alpha}{A}  T^{\hat r \hat 0}  
	= - \frac{\alpha}{4 \pi A} ({\bf E \times B})^{\hat r}.
\end{equation}
Here we have used the results of Section 7 of Paper I to evaluate the
source terms. The fields ${\bf E}$ and ${\bf B}$ can now be expressed
in terms of $e$ and $a$ using equations (\ref{Br}), (\ref{Btheta}) and
(\ref{Ephi}) for an $l=1$ dipole field. The angular integrals in
equation (\ref{AE}) and (\ref{AJ}) can be carried out analytically,
yielding
\begin{equation}
{\mathcal E}(t) = \frac{1}{2} \int_{r_1}^{r_2} dr (\alpha A^2) 
	\left( \frac{2}{3} A^3 e^2 + \frac{4}{3} \, \frac{a^2}{A^3 r^2} 
	+ \frac{2}{3} \, \frac{(a_{,r})^2}{A^3} \right)
\end{equation}
and 
\begin{equation}
{\mathcal J}(t) = \frac{2}{3} \alpha^2 A \int_{t_1}^{t_2} dt (a_{,r} e).
\end{equation}
These expressions can now be inserted into equation (\ref{Econs}) to
verify energy conservation in Schwarzschild coordinates.

\subsection{Kerr-Schild coordinates}
\label{secA.3}

In Kerr-Schild coordinates, the determinant of the metric is
\begin{equation}
\sqrt{-g} = r_s^2 \sin \theta.
\end{equation}
The fluxes $J^0$ and $J^{r_s}$ in (\ref{AE}) and (\ref{AJ}) are 
easiest evaluated by using the form (I.7.8) for the electromagnetic
stress-energy tensor $T^{ab}$.  We then find
\begin{equation}
4 \pi J^0 = 4 \pi T^0_{~0} = 
	\frac{1}{2} n^0 n_0 (E_i E^i + B_i B^i) + n^0 \epsilon_{0lm}
	E^l B^m.
\end{equation}
Here $\epsilon_{0lm} = - \beta \epsilon_{lr_s m}$ (see (I.3.19)) and
restricting the summations to the only non-vanishing components
$E^{\hat \phi}$, $B^{\hat r}$ and $B^{\hat \theta}$ yields
\begin{equation}
4 \pi J^0 = - \frac{1}{2} ({\bf E}^2 + {\bf B}^2) 
	- \frac{1 + 2M/r_s}{1 + r_s/{2M}} E^{\hat \phi} B^{\hat \theta}.
\end{equation}
The radial component can be evaluated in a similar way, which 
leads to
\begin{eqnarray}
4 \pi J^r & = & \frac{1}{1 + r_s/(2M)} \left( 
	(E^{\hat \phi})^2 + (B^{\hat \theta})^2\right) \\
& & 	+ \left( \frac{1 + 2M/r_s}{(1 + r_s/(2M))^2} 
	+ \frac{1}{1+2M/r_s} \right)
	E^{\hat \phi} B^{\hat \theta}. \nonumber
\end{eqnarray}
Specializing to $l=1$ dipole fields, the angular integrals in equation
(\ref{AE}) and (\ref{AJ}) can again be carried out 
analytically, which yields
\begin{eqnarray}
{\mathcal E}(t) & = & \frac{1}{2} \int_{r_1}^{r_2} dr_s r_s^2
	\Big( \frac{2}{3} \frac{e^2}{r_s^2} + 
	\frac{4}{3} \frac{a^2}{r_s^4} \\
& &	+ \frac{2}{3} \frac{(a_{,r_s})^2}{(1 + 2M/r_s) r_s^2}
	+ \frac{4}{3} \frac{e a_{,r_s}}{r_s^2} \, 
	\frac{(1+2M/r_s)^{1/2}}{1+r_s/(2M)} \Big) \nonumber
\end{eqnarray}
and
\begin{eqnarray}
{\mathcal J}(t) & = &  \frac{2}{3} \int_{t_t}^{t_2} dt \Big\{
	\frac{1}{1 + r_s/(2M)} 
	\Big( e^2 + \frac{(a_{,r_s})^2}{1+2M/r_s} \Big) \\
& & 	+ \Big(\frac{(1+2M/r_s)^{1/2}}{(1 + r_s/(2M))^2} +
	\frac{1}{(1+ 2M/r_s)^{3/2}} \Big) e a_{,r_s} \Big\} \nonumber
\end{eqnarray}
These expressions can be inserted into equation (\ref{Econs}) to
verify energy conservation in Kerr-Schild coordinates.

\section{Finite-Difference Equations}
\label{secB}

To finite difference the coupled evolution equations (\ref{edot1}) and
(\ref{adot1}) for Schwarzschild or maximal slicing 
(or (\ref{edot_ks}) and (\ref{adot_ks}) in the
Kerr-Schild case) we introduce a uniform radial grid $r_i, i = 1, 2,
... , I$.  We regrid the radial gridpoints after each
timestep, using linear interpolation between the old gridpoints, so
that the stellar surface $R_s$ is always half-way between the first two
gridpoints at $(r_1 + r_2)/2$.  In the Kerr-Schild simulations this is
no longer necessary once the interior of the black hole has been
excised and the inner boundary condition is imposed at a constant
excision radius $R_{\rm ex}$.  All variables are defined at the
gridpoint $r_i$.  Note that our numerical grid never extends to $r =
0$.  In the following we use $\Delta t = t^{n+1} - t^n$ and $\Delta r
= r_{i+1} - r_i$.

\subsection{Schwarzschild Coordinates and Maximal Slicing}
\label{secB1}

For isotropic Schwarzschild coordinates (Section \ref{sec7.2}) and
maximal slicing (Section \ref{sec7.3}) we adopt an implicit evolution
scheme.  It is convenient to use the abbreviations
\begin{equation}
B = \frac{\alpha}{A}
\end{equation}
and
\begin{equation}
C = A^3.
\end{equation}
Finite difference representations of equations (\ref{edot1}) and
(\ref{adot1}) can then be written as
\begin{equation} \label{adot_fd}
\frac{a^{n+1}_i - a^n_i}{\Delta t} = 
B^{n+1}_i C^{n+1}_i e^{n+1}_i + 
\beta^{n+1}_i \left( \frac{a^{n+1}_{i+i} - a^{n+1}_{i-1}}{2 \Delta r} \right)
\end{equation}
and 
\begin{eqnarray} \label{edot_fd}
& & \frac{e^{n+1}_i - e^n_i}{\Delta t} = 
\frac{1}{C^{n+1}_i} \, \frac{1}{(\Delta r)^2} \\
& &  
\times \left( B^{n+1}_{i + 1/2} (a^{n+1}_{i+1} - a^{n+1}_{i})
	-  B^{n+1}_{i - 1/2} (a^{n+1}_{i} - a^{n+1}_{i-1}) \right) \nonumber \\
& & - \frac{B^{n+1}_i}{C^{n+1}_i} \frac{l(l+1)}{r_i^2} a^{n+1}_i
	+ \beta^n_i \frac{r_i^2}{2 \Delta r} 
	\left(\frac{e^n_{i+1}}{r_{i+1}^2} - 
	\frac{e^n_{i-1}}{r_{i-1}^2} \right). \nonumber
\end{eqnarray}
Here the midpoint values $B^{n+1}_{i + 1/2}$ are computed as linear
interpolations $(B^{n+1}_{i + 1} + B^{n+1}_{i})/2$.  Note also that we
have finite-differenced the shift term implicitly in (\ref{adot_fd}),
but explicitly in (\ref{edot_fd}). 

The boundary condition (\ref{a_inner_condition}) is implemented as
\begin{equation} \label{a_ib}
\frac{a^{n+1}_1 + a^{n+1}_2}{2} = - (A^{n+1}_{3/2} r_{3/2})^2 \frac{B}{2},
\end{equation}
where the interior field strength $B$ is given by equation
(\ref{Bbc2}).  The outer boundary condition (\ref{outer_boundary})
is finite differenced as
\begin{equation} \label{a_ob}
\frac{a^{n+1}_{I} - a^{n}_{I}}{\Delta t}
= - \frac{a^{n}_{I} - a^{n}_{I-1}}{\Delta r}
	- \frac{a^n_{I}}{r_{I}}
\end{equation}

Substituting the finite difference equation (\ref{edot_fd}) into
(\ref{adot_fd}) and using equations (\ref{a_ib}) and (\ref{a_ob}) at
the inner and outer boundaries yields a single tridiagonal matrix
equation for $a^{n+1}_i$.  For this approach to work we had to
finite-difference the shift terms in (\ref{edot_fd}) explicitly,
because these would have otherwise introduced $e^{n+1}_i$ terms into
the equation for $a^{n+1}_i$.  We invert the tridiagonal matrix at
each time step to obtain the new values $a^{n+1}_i$.  New values
$e^{n+1}_i$ can then be found by using equation (\ref{edot_fd}) again
together with the inner boundary condition (\ref{e_inner_condition})
\begin{equation}
\frac{e^{n+1}_1 + e^{n+1}_2}{2} = 
	\frac{v^{n+1}_{3/2}}{C^{n+1}_{3/2}} \, 
	\frac{a^{n+1}_2 - a^{n+1}_1}{\Delta r}
\end{equation}
and an outer boundary condition equivalent to (\ref{a_ob}).

To choose the size for successive timesteps we set
\begin{equation} \label{Cour}
\Delta t = {\mathcal C} ~{\rm min} \left(\Delta r \right),
\end{equation}
where we have introduced a Courant factor ${\mathcal C}$ in the above 
equation. Implicit differencing 
imposes no restrictions on ${\mathcal C}$, but accuracy usually requires
the choice ${\mathcal C} \lesssim 1$. 

\subsection{Kerr-Schild Slicing}
\label{secB2}

The evolution equations (\ref{edot_ks}) and (\ref{adot_ks}) in
Kerr-Schild slicing can be finite differenced in a very similar
fashion to those in Schwarzschild coordinates or maximal slicing.

In addition to an implicit scheme, we also implemented
an iterative explicit scheme, in which case the finite difference
equations are written as
\begin{equation} \label{adot_fdks}
\frac{a^{n+1}_i - a^n_i}{\Delta t} = \alpha^{n}_i e^{n}_i +
\beta^{n}_i \left( \frac{a^{n}_{i+i} - a^{n}_{i-1}}{2 \Delta r}
\right)
\end{equation}
and
\begin{eqnarray} \label{edot_fdks}
& & \frac{e^{n+1}_i - e^n_i}{\Delta t} = 
\frac{\alpha^{n}_i}{(\Delta r)^2} \\
& &  \times \left(
(\alpha^{n}_{i+1/2})^2(a^{n}_{i+1} - a^{n}_{i}) -
(\alpha^{n}_{i-1/2})^2(a^{n}_{i} - a^{n}_{i-1}) \right) \nonumber\\
& & - \alpha^{n}_i \frac{l(l+1)}{r_i^2} a^{n}_i 
	+ \alpha^{n}_i K^{n}_i e^{n}_i\nonumber \\
& & + \beta^{n}_i \frac{r_i^2}{2 \Delta r}
	\left(\frac{e^n_{i+1}}{r_{i+1}^2} - 
	\frac{e^n_{i-1}}{r_{i-1}^2} \right). \nonumber
\end{eqnarray}
Here the radius $r$ denotes an areal radius in this Section (denoted
with $r_s$ in the rest of the paper).  Prior to excision, the boundary
conditions (\ref{a_inner_ks}) and (\ref{e_inner_ks}) can be
finite-differenced as
\begin{equation}
\frac{a^{n+1}_1 + a^{n+1}_2}{2} = - \frac{(r_{3/2})^2 B}{2}
\end{equation}
and
\begin{equation}
\frac{e^{n+1}_1 + e^{n+1}_2}{2} = 
	\frac{v^{n+1}_{3/2}}{(1+2M/r_{3/2})^{1/2}} \, 
	\frac{a^{n+1}_2 - a^{n+1}_1}{\Delta r}
\end{equation}
Once the stellar radius is smaller than the excision radius $R_{\rm ex}$
we switch to the excision boundary conditions (\ref{inner_ex}), which
can be implemented as
\begin{eqnarray} 
a^{n+1}_1 & = & a^{n+1}_2 \label{B_inner_bc_a}\\
e^{n+1}_1 & = & e^{n+1}_2 \label{B_inner_bc_b}
\end{eqnarray}
The outer boundaries are identical to those for Schwarzschild
coordinates and maximal slicing.  

An iterative Crank-Nicholson scheme can be constructed in the
following way (cf.~Teukolsky 2000).  In a predictor step, the above
equations are solved for approximate new values $\tilde{a}^{n+1}_i$
and $\tilde{e}^{n+1}_i$.  The average between these values and the old
values ${a}^{n}_i$ and ${e}^{n}_i$ is then used on the right hand side
to compute improved new values $\bar{a}^{n+1}_i$ and $\bar{e}^{n+1}_i$
in a first corrector step.  These are again averaged with ${a}^{n}_i$
and ${e}^{n}_i$ and used on the right hand sides for the second and
final corrector step, which yields the new values ${a}^{n+1}_i$ and
${e}^{n+1}_i$.

The Kerr-Schild results presented here were obtained with the above explicit
scheme.  However, we also implemented an implicit scheme very similar
to that in Section \ref{secB1} and obtained very similar results.  In
particular, we found the interesting result that excision boundary conditions
(\ref{inner_ex}) can be used with an implicit scheme without encountering 
problems.  Implemented as in Section \ref{secB1}, the implicit scheme is first
order in time (but could easily be made second order), while the above
iterative scheme is second order in time, which resulted in more
accurate results in late-time evolutions.

\end{appendix}

\end{document}